\documentclass[journal]{IEEEtran}
\usepackage{graphicx}   
\usepackage{tikz}
\usetikzlibrary{shapes.geometric, arrows.meta, positioning, calc}
\usepackage{float}
\usepackage{pgfplots}    
\pgfplotsset{compat=1.18} 
\usepackage{amsmath}    
\usepackage{circuitikz}
\usepackage{booktabs}   
\usepackage{multirow}   
\usepackage{tabularx}
\usepackage[
style=numeric, 
backend=biber, 
labelnumber=true,
sorting=nyt
]{biblatex}
\usepackage[hidelinks]{hyperref}
\usepackage[autostyle]{csquotes}
\MakeOuterQuote{"}
\begin{document}

	\title{Algorithmic Energy Management in Constrained Railway Traction Networks: A Systematic Review} 
	
	\author{Márton~László~Ambrus,
        Stuart Hillmansen,
        and Zhongbei Tian
\thanks{M. L. Ambrus, S. Hillmansen, and Z. Tian are with the Dept. of Electronic, Electrical and Systems Engineering, University of Birmingham, BCRRE, Birmingham, United Kingdom (e-mail: m.l.ambrus@bham.ac.uk; s.hillmansen@bham.ac.uk; z.tian@bham.ac.uk).}}
	
	\maketitle

	\begin{abstract}
		The decarbonisation of heavy-duty railway networks requires maximising the capacity of existing electrical infrastructure. Integrating heavy freight alongside fast passenger services exposes the hard physical limits of conventional alternating current traction networks, causing severe localised power quality degradation, phase unbalance, and low-voltage behaviour that triggers protective substation tripping. Because upgrading physical hardware is highly capital-intensive, software-based Energy Management Strategies (EMS) offer a potentially viable alternative for preventing these power capacity challenges. This systematic review synthesises the literature on algorithmic energy management for grid-constrained multi-train AC railway networks, classifying the reviewed studies along three axes: algorithm family, operational scope, and constraint coupling. The review documents three consistent findings across the included studies. First, single-train trajectory optimisation, however mathematically refined, cannot represent the coupled electrical interactions that increasingly define network capacity on mixed-traffic networks. Second, while multi-train Train-Track-Power (TTP) simulations correctly capture these interactions, the algorithm families currently used to solve them face well-documented trade-offs between computational tractability and constraint flexibility; predictive and distributed methods, including hierarchical model predictive control and decomposition-based schemes, narrow this trade-off substantially, but the review finds that electrical fidelity and network-scale real-time operation have been demonstrated largely in separate studies rather than together. Third, the literature increasingly identifies a gap between mathematically optimal speed profiles and operationally executable ones, particularly for networks operated by human drivers rather than Automatic Train Operation systems. The review delineates where current methods succeed, where they fail, and which directions the literature has identified as open.
	\end{abstract}
	
	
	\section{Drivers, Constraints and the Role of Energy Management}
	
	\subsection{The Policy Context for Network Efficiency}
	
	Historically, maximising the efficiency of railway electrification has been driven by fundamental engineering best practices and the need to reduce operational energy costs. Improving traction efficiency and reducing transmission losses have always been inherent goals of railway engineers. However, the transition toward zero-emission heavy-duty transport is now being accelerated by international climate policies and legally binding carbon budgets.
	
	National policy frameworks formalise this acceleration without addressing the technical capacity of the physical grid required to absorb the resulting electrical load. At European level, the \textit{Sustainable and Smart Mobility Strategy} \cite{ec2020mobility} sets targets to double high-speed rail traffic by 2030 and triple rail freight by 2050, while UK strategic frameworks such as \cite{decarbonising_uk_transport} and \cite{departmentfortransport} establish strict legislative milestones, most notably the phase-out of diesel rail traction by 2040. Crucially, these roadmaps explicitly exclude "changing travel behaviour" (demand reduction) from their scope, assuming that technological substitution alone must achieve Net Zero targets. Consequently, if logistical demand remains constant while traditional high-energy-density diesel is banned, operators are forced to extract maximum capacity from existing, heavily strained electrical infrastructure.

	The scale of the residual electrification task, and therefore of the load to be absorbed, varies widely between networks. UIC statistics \cite{uic2025statistics} report Switzerland as effectively fully electrified, with India having electrified approximately 99\% of its broad-gauge network, followed by China (82\%), Spain (67\%), Japan (64\%), France (60\%), Russia (52\%) and the United Kingdom (39\%). These figures describe route-kilometres rather than traffic: as the IEA notes \cite{iea2019rail}, the share of traffic hauled electrically is typically far higher than the share of network electrified, because electrification is concentrated on the busiest corridors. The consequence is that the corridors carrying the densest mixed traffic are precisely those where the traction supply is already most heavily loaded, and this holds across national contexts rather than being a peculiarity of any single network. The constraint problem examined in this review is therefore general, although the illustrative infrastructure examples used throughout are drawn predominantly from 25\,kV 50\,Hz mainline practice.
	
	\subsection{The Limitations of Full Electrification and the "Power Headroom" Deficit}
	
	The default technological solution for decarbonising heavy-duty rail transport has been direct overhead line electrification (OLE). While OLE provides highly efficient energy transfer, extending catenary infrastructure across all long-distance freight and regional networks is fundamentally constrained by prohibitive capital costs, geographical barriers, and grid capacity limitations.

	This is a constraint on cost and pace rather than on feasibility. Switzerland has achieved effectively complete electrification of its national network \cite{uic2025statistics} and is therefore the natural reference case for the fully electrified end state that decarbonisation policies aim to achieve. That case is instructive precisely because it does not solve the problem examined here: where electrification is complete, every service becomes a catenary load, so the aggregate demand placed on the traction supply is higher rather than lower, and the ability to determine the power headroom available within a given feeding section becomes more valuable, not less. The Swiss network is supplied at 15\,kV 16.7\,Hz from a dedicated traction transmission system rather than at 25\,kV 50\,Hz; the coverage argument made here is independent of that distinction, whereas the electrical detail developed in Sections \ref{electrification_systems} and \ref{sec:power_quality} is specific to 25\,kV 50\,Hz practice.
	
	Beyond the static costs of infrastructure, a dynamic operational gap exists in the current electrified network model: the inability to accurately simulate and predict the real-time power headroom available to trains moving along a railway section. As heavy-duty trains travel across varying topographies, encountering steep gradients causes sudden, significant spikes in power demand. Electrical grid supply models currently struggle to provide reliable, localised headroom predictions for these dynamic moving loads. If a fully electric heavy-duty train enters a track section with insufficient grid headroom during a high-power demand event, the system risks tripping the power supply, or the train is forced to drastically limit its speed. This lack of predictable power makes relying solely on hardware-side electrification an unacceptable risk for high-capacity mixed-traffic operation, motivating software-side energy management strategies as the subject of this review.

	\subsection{The Role of Energy Management Strategies (EMS)}
	\label{sec:ems_role}
	
	Having established that the physical infrastructure of AC traction networks is constrained by low-voltage behaviour, phase unbalance, and the continuous load profiles of heavy freight, the literature demonstrates a clear consensus that hardware upgrades alone are economically prohibitive and insufficient to solve these challenges.
	
	Consequently, network operators must rely on software-based Energy Management Strategies (EMS) to dynamically allocate power and manage onboard energy buffers. However, the literature shows a significant debate regarding the optimal architectural scope of these systems.
	
	\subsubsection{The Limitations of Static Rule-Based Control}
	
	Rule-based strategies, implemented as state machines, thermostat-style buffer control or fuzzy-logic supervisors, form a well-established baseline for \textit{onboard} energy management in hybrid and fuel-cell rail vehicles \cite{peng2020rulebased, deng2021adaptive}. They are causal, computationally trivial, and straightforward to certify. Notably, \cite{peng2020rulebased} derives its rule set directly from the results of dynamic programming and reports fuel economy close to the dynamic-programming optimum on regional duty cycles, which establishes that under \textit{known and repeatable} driving cycles a well-tuned rule base can approach optimality rather than being categorically inferior to it.

	The limitation lies elsewhere. Because rule-based logic is reactive by construction, it conditions only on the vehicle's present state and cannot anticipate an upcoming gradient or a developing constraint elsewhere on the feeding section. This matters little when the duty cycle is known in advance and the electrical boundary condition is fixed, and a great deal when neither holds. The literature has therefore shifted toward predictive formulations for precisely those cases in which route topography and catenary voltage limits must be treated as forward-looking, multi-objective constraints. It should be stressed that this argument concerns the reactive-versus-predictive distinction; it is not a claim that rule-based logic constitutes the prevailing architecture for \textit{grid-side} energy management, where, as Section \ref{sec:predictive} discusses, optimisation-based supervisory schemes are already well represented.
	
	\subsubsection{The Shift from Single-Train to Multi-Train Simulation}
	
	Earlier research \cite{energy_efficient_train_control, single_train_trajectory_optimisation} focused mainly on single-train trajectory optimisation. These studies successfully derived the theoretical limits of mechanical energy consumption for an individual vehicle, typically by optimising the sequence of maximum acceleration, cruising, coasting, and maximum braking phases.
	
	However, more recent literature argues that this approach is limited for heavy-duty applications because it operates under a "grid-blind" assumption: single-train models treat the traction network as an infinite power source. Because such models do not account for the presence of other vehicles within the same electrical sector, they cannot predict or prevent the simultaneous traction overloads discussed in Section \ref{sec:power_quality}. A grid-blind simulation may generate a theoretically optimal eco-driving profile that mathematically dictates maximum acceleration for a passenger train at the exact moment a heavy freight train is dragging the network voltage down, rendering the profile operationally hazardous.
	
	\subsubsection{Multi-Train Simulation and the Computational Bottleneck}
	
	To resolve the grid-blind limitations of single-train models, recent literature has called for a shift toward Multi-Train Comprehensive Optimisation (MTCO). Authors such as \cite{tao_2024_comprehensive, multipletrain_trajectory_optimisation} argue that by simulating the Train-Track-Power (TTP) system simultaneously, network-level EMS can actively stagger the acceleration phases of heavy freight and passenger trains, flattening the total instantaneous power demand.
	
	This peak shaving ensures the network does not exceed its Firm Service Capacity (FSC). Because overstepping the FSC incurs severe financial penalties for the railway operator, the literature emphasises that dynamic load shifting is not merely a technical requirement to respect hardware limits, but also a primary economic objective. Furthermore, studies show MTCO allows for the active synchronisation of regenerative braking between vehicles, further reducing grid power draw.
	
	However, while researchers broadly agree that multi-train simulation represents the necessary direction, a critical tension emerges in the literature regarding its implementation. As the EMS must calculate the simultaneous speed profiles, locations, and electrical interactions of several trains across a wide range of track segments, the state-space of the simulation grows exponentially. This computational bottleneck forces a recurring methodological choice: the algorithmic architectures reviewed in Sections \ref{sec:limitations}--\ref{sec:predictive} consistently trade mathematical optimality against computational viability, and the literature offers no single dominant solution.
	
	\subsection{Scope and Research Questions of this Review}
	\label{sec:scope}
	
	This review systematically synthesises the literature on algorithmic energy management strategies for AC-electrified railway networks operating under multi-train, grid-constrained conditions. Specifically, it addresses three research questions: (i) how do existing algorithm families handle the coupling between train trajectory optimisation and traction power supply constraints; (ii) what trade-offs exist between solution optimality, computational tractability, and operational deployability; and (iii) what classification of approaches emerges from the literature, and where are the unresolved gaps?
	
	The remainder of this paper is organised as follows. Section \ref{sec:ems_role} outlines the role of energy management strategies. Section \ref{sec:methodology} describes the PRISMA-based review methodology and introduces the classification framework used to organise the reviewed literature. Sections \ref{electrification_systems}--\ref{sec:mixed_fleet} establish the physical infrastructure constraints and the mixed-fleet operational context that motivate algorithmic intervention. Sections \ref{sec:limitations}, \ref{sec:heuristics} and \ref{sec:predictive} review the three principal algorithm families found in the literature: analytical/exact methods, meta-heuristic methods, and predictive/distributed methods, respectively. Section \ref{sec:topography} reviews how topographical constraints are integrated into these algorithms. Section \ref{sec:synthesis} synthesises the comparative trade-offs across the reviewed literature, and Section \ref{sec:gaps} identifies the open research directions the literature points toward.

	\section{Review Methodology}
	\label{sec:methodology}
	
	To ensure a comprehensive, transparent, and reproducible synthesis of the literature on algorithmic energy management strategies for grid-constrained multi-train AC railway networks, this review followed the Preferred Reporting Items for Systematic Reviews and Meta-Analyses (PRISMA) framework. The full PRISMA selection workflow used in this review is summarised in Fig. \ref{fig:prisma}.

	\subsection{Classification Framework Used in this Review}
	\label{sec:taxonomy}
	
	To structure the synthesis of the reviewed literature, this review classifies each included study along three axes:
	\begin{itemize}
		\item \textbf{Axis A --- Algorithm family:} Analytical methods (Pontryagin's Maximum Principle, Calculus of Variations); exact discrete methods (Dynamic Programming, Mixed-Integer Linear Programming, Brute Force); meta-heuristic methods (Genetic Algorithms, Particle Swarm Optimisation, Ant Colony Optimisation, NSGA-II); predictive and distributed methods (Model Predictive Control, hierarchical and distributed MPC, rolling-horizon control, and decomposition schemes including Lagrangian relaxation and the alternating direction method of multipliers); and learning-based methods.
		\item \textbf{Axis B --- Operational scope:} Single-train trajectory optimisation; multi-train timetable-only optimisation; and multi-train Train-Track-Power (TTP) coupled optimisation.
		\item \textbf{Axis C --- Constraint coupling:} Mechanical constraints only (mass, gradient, speed limits); mechanical plus electrical constraints (substation power, catenary voltage, FSC); and mechanical plus electrical plus stochastic constraints (passenger load, demand uncertainty, delay propagation).
	\end{itemize}
	
	Sections \ref{sec:limitations}, \ref{sec:heuristics} and \ref{sec:predictive} are organised around Axis A, mapping each reviewed algorithm family onto its representative scope (Axis B) and the constraints it has been demonstrated to handle (Axis C). Tables \ref{tab:algorithm_evolution} and \ref{tab:execution_times} present the structured outcome of this classification.
	
	\subsection{Information Sources and Search Strategy}
	
	The primary literature search was conducted across leading engineering, scientific, and multidisciplinary databases: \textit{ScienceDirect (Elsevier)}, \textit{IEEE Xplore}, and \textit{Google Scholar}.
	
	To capture the intersection of vehicle physics, high-voltage electrical infrastructure, and advanced mathematical solvers, the search strategy utilised Boolean operators (AND/OR). Keywords were combined across four core thematic domains to ensure both physical hardware limits and software solutions were comprehensively reviewed:
	
	\begin{itemize}
		\item \textbf{Context \& Operations:} \texttt{"Railway Transportation" OR "Heterogeneous Traffic" OR "Mixed Fleet" AND "Decarbonisation" OR "Firm Service Capacity"}
		\item \textbf{Electrical Infrastructure \& Power Quality:} \texttt{"Static Frequency Converters" OR "SFC" OR "Power Quality" OR "Negative Sequence Components" OR "Low Voltage Behaviour" AND "AC Traction Networks"}
		\item \textbf{Simulation Architectures:} \texttt{"Multi-train" OR "Single-train" AND "Train-Track-Power" OR "TPSS" AND "Peak Shaving" OR "Regenerative Braking Synchronisation"}
		\item \textbf{Algorithms \& Control:} \texttt{"Energy Management Systems" AND "Genetic Algorithms" OR "Dynamic Programming" OR "Reinforcement Learning" AND "Computational Complexity" OR "Curse of Dimensionality"}
	\end{itemize}
	
	Due to the multi-disciplinary nature of this review, which bridges electrical engineering, railway operations, and algorithm design, a hybrid search strategy was employed. In addition to primary database keyword searches, \textit{Backward Citation Searching} was used. The reference lists of foundational, highly relevant papers identified in the initial search were systematically screened to discover further literature regarding heuristic solvers, hardware constraints, and power quality limitations that primary database queries may have excluded.
	
	\subsection{Eligibility Criteria}
	
	To filter the initial search results and isolate papers directly relevant to the operational control of heavy-duty fleets and their electrical infrastructure, strict inclusion and exclusion criteria were applied.
	
	\textbf{Inclusion Criteria:}
	\begin{enumerate}
		\item Peer-reviewed journal articles, international conference proceedings, and foundational academic theses.
		\item Studies explicitly modelling advanced traction systems, multi-train simulations (e.g., Train-Track-Power models), energy management, or trajectory optimisation for heavy-duty, freight, or mainline rail networks.
		\item Research addressing physical vehicle constraints (e.g., heterogeneous mass, topography) or high-voltage infrastructure constraints (e.g., Firm Service Capacity, substation limits, low-voltage behaviour).
		\item Publications written in English.
	\end{enumerate}
	
	\textbf{Exclusion Criteria:}
	\begin{enumerate}
		\item Studies exclusively focusing on light-duty passenger road vehicles or dedicated urban metros (light-rail) without scalable applications to heavy-duty, mixed-traffic environments.
		\item Studies of hydrogen production, supply chain, or material science lacking direct application to vehicle-level or grid-level algorithmic control.
		\item Outdated computational studies (prior to 2000), unless established as foundational mathematical benchmarks (e.g., core proofs of Pontryagin's Maximum Principle or classic Dynamic Programming).
	\end{enumerate}

	Learning-based methods were included in both the search strategy set out above and the Axis A classification. No study applying reinforcement learning, learned models or other data-driven control to grid-coupled multi-train operation met the eligibility criteria, and the family is therefore represented in the classification scheme but unpopulated in the reviewed corpus. Its absence is a finding of the search rather than a restriction of its scope, and the open questions it raises are discussed in Section \ref{sec:gaps}.
	
	\subsection{Selection Process and PRISMA Framework}
	
	The filtering and selection process was conducted in sequential phases, as illustrated in Fig. \ref{fig:prisma}. Following the execution of the search strings across the selected databases, duplicate records were identified and removed. Initial screening was conducted by evaluating the titles and abstracts of the remaining records to determine their relevance to mixed-mode traction, electrical grid constraints, and algorithm design.
	
	Literature that passed the initial screening was reviewed in full. During this phase, papers were assessed regarding their mathematical rigor, the complexity of the physical constraints handled (e.g., Firm Service Capacity, low-voltage limits), and their applicability to infrastructure-aware or mixed-traffic fleet architectures. Following this filtering process, a corpus of 44 studies was identified for in-depth synthesis at the time of initial submission. During revision, a supplementary search targeting predictive, rolling-horizon and distributed methods was conducted using the
	Algorithms and Control search string extended with the terms "Model Predictive Control", "rolling horizon", "receding horizon", "decomposition" and "distributed
	optimisation". This search returned 18 records, of which 11 met the eligibility criteria and were added to the corpus, giving a final total of 55 included studies.
	In Fig. \ref{fig:prisma} the supplementary search is reported alongside backward citation searching under identification via other methods, the two sources together contributing 35 of the 55 included studies.
	
	In accordance with PRISMA 2020 \cite{page2021prisma, page2021prismaee}, exclusions at the title-and-abstract stage are reported in aggregate, whereas exclusions at the full-text stage are itemised by reason in Fig. \ref{fig:prisma}. Of the 15 reports excluded after full-text assessment, exclusions were attributable to: wrong setting, being confined to light-duty road vehicles or dedicated urban metro systems without scalable application to heavy-duty mixed traffic ($n=6$); wrong intervention, addressing hydrogen production, supply chain or material science without vehicle-level or grid-level algorithmic control ($n=4$); wrong outcome, reporting neither an optimisation method nor a quantified energy or electrical-constraint outcome ($n=3$); and full text not retrievable within the review period ($n=2$). The terminology of records, reports and studies used in Fig. \ref{fig:prisma} follows the definitions given in \cite{page2021prisma}, and the reporting of the search strategy in the preceding subsections follows the PRISMA-S extension \cite{rethlefsen2021prismas}.
	
	\begin{figure*}[t]
	\centering
	\resizebox{\textwidth}{!}{
		\begin{tikzpicture}[
			node distance=0.8cm and 0.5cm,
			mainbox/.style={rectangle, draw, minimum width=4.8cm, minimum height=1.5cm, text width=4.5cm, align=left, font=\small},
			sidebox/.style={rectangle, draw, minimum width=3.8cm, minimum height=1cm, text width=3.5cm, align=left, font=\small},
			header/.style={rectangle, draw, fill=yellow!60, minimum width=9.1cm, minimum height=0.6cm, font=\bfseries\small, align=center},
			phase/.style={rectangle, draw, fill=blue!20, rounded corners=1mm, font=\bfseries\small, align=center},
			arrow/.style={-{Stealth[scale=1.2]}, thick}
			]

			\node[mainbox] (id1) {Records identified from\\ Databases ScienceDirect, IEEE\\ Xplore, Google Scholar ($n = 120$)};

			\node[mainbox, below=1.0cm of id1] (screen1) {Records screened\\ ($n = 120$)};
			\node[sidebox, right=0.5cm of screen1] (ex1) {Records excluded\\ ($n = 85$)};

			\node[mainbox, below=1.0cm of screen1] (retrieve1) {Reports sought for retrieval\\ ($n = 35$)};
			\node[sidebox, right=0.5cm of retrieve1, text width=4.6cm, minimum width=4.9cm] (ex2) {Reports excluded ($n = 15$):\\ Wrong setting: light-duty road or\\ dedicated urban metro ($n = 6$)\\ Wrong intervention: hydrogen\\ production / materials ($n = 4$)\\ Wrong outcome: no method or\\ quantified outcome ($n = 3$)\\ Full text not retrieved ($n = 2$)};

			\node[mainbox, below=1.9cm of retrieve1] (inc1) {Studies included from\\ databases ($n = 20$)};

			\node[mainbox, right=5.9cm of id1, text width=4.5cm] (id2) {Records identified from:\\[1mm] Citation searching ($n = 35$)\\ Supplementary search at\\ revision ($n = 18$)};

			\node[mainbox, at={(id2 |- screen1)}] (screen2) {Records screened\\ ($n = 53$)};
			\node[sidebox, right=0.5cm of screen2] (ex3) {Records excluded\\ ($n = 18$)};

			\node[mainbox, at={(id2 |- inc1)}] (inc2) {Studies included from\\ other methods ($n = 35$)};

			\node[mainbox, below=1.6cm of inc1] (final) {Studies included in review\\ ($n = 55$)};

			\node[header, above=0.6cm of id1.north west, anchor=south west] (head1) {Identification of studies via databases and registers};
			\node[header, above=0.6cm of id2.north west, anchor=south west] (head2) {Identification of studies via other methods};

			\path (screen1.north west) -- (screen1.west |- inc1.south) coordinate[midway] (mid_screen_left);
			\node[phase, minimum width=7.4cm, rotate=90] (phase2) at ($ (mid_screen_left) + (-1.2cm, 0) $) {Screening};

			\node[phase, minimum width=2.8cm, rotate=90] (phase1) at ($ (id1.west) + (-1.2cm, 0) $) {Identification};

			\node[phase, minimum width=2cm, rotate=90] (phase3) at ($ (final.west) + (-1.2cm, 0) $) {Included};

			\draw[arrow] (id1) -- (screen1);
			\draw[arrow] (screen1) -- (ex1);
			\draw[arrow] (screen1) -- (retrieve1);
			\draw[arrow] (retrieve1) -- (ex2);
			\draw[arrow] (retrieve1) -- (inc1);
			\draw[arrow] (inc1) -- (final);

			\draw[arrow] (id2) -- (screen2);
			\draw[arrow] (screen2) -- (ex3);
			\draw[arrow] (screen2) -- (inc2);

			\draw[arrow] (inc2.south) |- (final.east);

		\end{tikzpicture}
	}
	\caption{PRISMA 2020 flow diagram detailing the literature search and selection process. Studies were identified through database querying, backward citation searching, and a supplementary search conducted at revision stage covering predictive, rolling-horizon and distributed methods; the latter two sources are reported together under identification via other methods. Following \cite{page2021prisma}, title-and-abstract exclusions are reported in aggregate while full-text exclusions are itemised by reason. The 55 included studies constitute the reviewed corpus; the reference list additionally contains standards, policy and statistical sources, and methodological references cited as supporting material, which were not identified through the search strategy and are therefore not counted here.}
	\label{fig:prisma}
\end{figure*}
	
	\subsection{Data Extraction and Synthesis}
	
	For each included study, the following parameters were extracted into a structured matrix: (i) algorithm family per Axis A of the classification framework introduced in Section \ref{sec:taxonomy}; (ii) operational scope per Axis B; (iii) constraint coupling per Axis C; (iv) reported computational latency, where stated by the original authors; and (v) explicitly stated limitations or gaps identified by the authors themselves. The full structured matrix is summarised in Tables \ref{tab:algorithm_evolution}, \ref{tab:topography_constraints}, and \ref{tab:execution_times}, which together form the evidentiary basis for the thematic synthesis presented in Sections \ref{sec:limitations}--\ref{sec:gaps} of this review.
	
	\section{Railway Electrification Systems} \label{electrification_systems}
	
	To understand the necessity of algorithmic power optimisation, it is first essential to establish the physical hardware constraints of the railway traction network. Modern high-speed and heavy-freight networks rely on infrastructure that must constantly balance the massive power demands of moving trains against the stability of the national electrical grid.
	
	\subsection{AC Traction Networks and Auto-Transformer Architectures}
	
	The standard configuration for modern electrified mainline railways, including the UK network, utilises a 25kV, 50Hz single-phase alternating current (AC) supply drawn directly from the national three-phase transmission grid. The permissible operating envelope for this system is defined by BS EN 50163 \cite{en50163} and its international equivalent IEC 60850 \cite{iec60850}, which specify a nominal voltage $U_n = 25$\,kV, a highest permanent voltage $U_{max1} = 27.5$\,kV, a highest non-permanent voltage $U_{max2} = 29$\,kV, a lowest permanent voltage $U_{min1} = 19$\,kV, and a lowest non-permanent voltage $U_{min2} = 17.5$\,kV. Under normal operating conditions the pantograph voltage must lie within $U_{min1} \leq U \leq U_{max2}$; excursions into the range $U_{min2} \leq U < U_{min1}$ are permitted only for limited durations. These values, rather than any per-unit convention, are the reference points used throughout this review.

	Because trains act as large, moving electrical loads, the voltage along the overhead line equipment (OLE) naturally drops the further a train travels from a feeder substation. To mitigate these significant voltage drops over long distances, network operators deploy Auto-Transformer (AT) feeding arrangements (Fig. \ref{fig:at_schematic}). AT systems effectively double the transmission voltage to 50kV (using a +25kV contact wire and a -25kV along-track feeder) while delivering the standard 25kV to the train itself, thereby reducing transmission losses and allowing substations to be spaced further apart.

	A point of terminology warrants clarification, since the same AT topology is designated differently across national practices. In GB and much of Western European practice the arrangement is termed "$2\times25$\,kV", referencing the nominal $U_n$ of each of the two 25\,kV sections either side of the rail reference. In Chinese and some continental practice the equivalent topology is termed "$2\times27.5$\,kV", referencing instead the highest permanent voltage $U_{max1}$ of 27.5\,kV specified in \cite{en50163}. That choice reflects operating practice rather than a different system: because pantograph voltage falls with distance from the feeding point, substation busbars are commonly energised above $U_n$, at or close to $U_{max1}$, precisely so that trains at the far end of a feeding section remain within the permitted range. The "$2\times27.5$\,kV" designation therefore names the voltage impressed at the substation, whereas "$2\times25$\,kV" names the nominal system voltage defined in \cite{en50163}. The two designations describe the same physical arrangement, differing only in whether the nominal or the highest permanent voltage is used as the naming reference. This review adopts the "$2\times25$\,kV" convention throughout, consistent with the nominal-voltage designation of \cite{en50163}, while noting that readers accustomed to the "$2\times27.5$\,kV" convention should read the two as equivalent.
	
	\begin{figure}[t]
		\centering
		\resizebox{\columnwidth}{!}{
			\begin{circuitikz}

				\draw[thick] (0,2) -- (7,2) node[right] {+25kV (OLE)};
				
				\draw[thick, dashed] (2,0) -- (7,0) node[right, fill=white, inner sep=2pt] {0V (Rails)}; 
				\draw[thick] (0,-2) -- (7,-2) node[right] {-25kV (Feeder)};

				\draw (0,-2) to[sV] (0,2);
				
				\node[left, align=right] at (-0.5, 0) {50kV\\Supply};

				\draw (2,2) to[L, l=AT, *-*] (2,0);
				\draw (2,0) to[L, *-*] (2,-2);
				\draw (2,0) -- (1.2,0) node[ground] {}; 
				
				\node[above, font=\bfseries] at (1, 2.7) {Feeder Substation};

				\draw (4.5,2) to[generic, l=Train, *-*] (4.5,0);
				
				\draw[->, thick, blue] (4, 2.3) -- (5, 2.3) node[right] {$v(t)$};

				\draw (7,2) to[L, l=AT, *-*] (7,0);
				\draw (7,0) to[L, *-*] (7,-2);
				
				\draw (7,0) -- (6.2,0) node[ground] {}; 
				\node[above, font=\bfseries] at (7, 2.7) {Trackside AT};
				
			\end{circuitikz}
		}
		\caption{Simplified architecture of a $2 \times 25$kV Auto-Transformer (AT) traction power network, illustrating a single 50kV source split into +25kV and -25kV relative to the running rails.}
		\label{fig:at_schematic}
	\end{figure}
	
	\subsection{Grid Integration, SFCs, and Co-Phase Architectures}
	
	While traditional transformer-based substations successfully step down voltage, drawing massive single-phase loads from a three-phase national grid introduces severe power quality issues, most notably phase unbalance, reactive power burdens, and harmonic distortion. To actively manage grid integration, operators are increasingly transitioning to Static Frequency Converters (SFCs) and advanced converter-based architectures. Fig. \ref{fig:sfc_comparison} contrasts the grid-side behaviour of these two architectures.

	The winding configuration of the conventional arrangement shown on the left of Fig. \ref{fig:sfc_comparison} merits precise description, as it is the source of the unbalance the SFC architecture is designed to eliminate. In a conventional single-phase traction transformer the primary is connected across \textit{two} of the three grid phases in a line-to-line configuration; the third phase is simply not used at that substation. On the secondary side, one terminal supplies the traction busbar and the catenary while the other is bonded to the running rails and the earthing network \cite{tu2019realtime}. The rail connection is therefore a property of the secondary return path, not a connection of the unused third grid phase. Because each substation loads only two phases, adjacent substations are conventionally supplied with rotated phase pairs so that the unbalance is distributed across all three phases of the transmission network in aggregate, even though it remains severe at any individual substation.

	Balanced special connections, by contrast, do use all three grid phases. As catalogued by \cite{tu2019realtime}, V/v, Scott, YNd11, YNvd and impedance-matching configurations convert the three-phase supply into one or two single-phase outputs specifically to mitigate negative-sequence unbalance, and the choice between them is largely a national and historical one: GB practice has favoured single-phase and V connections, while V/v and balanced connections are widespread in Chinese high-speed practice. Fig. \ref{fig:sfc_comparison} depicts the conventional two-phase case, which remains the dominant legacy arrangement on mixed-traffic mainline networks and is consequently the configuration against which the algorithmic mitigation strategies reviewed here must be assessed.
	
	Unlike standard mechanical transformers, SFCs utilise advanced power electronics to fully decouple the railway traction network from the national grid \cite{bowdidge_2023_a, ronanki_2018_modular}. Because SFCs draw a balanced load across all three phases, they can be safely connected to lower-voltage distribution networks (e.g., 33-66 kV) rather than requiring costly high-voltage transmission connections. This adaptability allows SFCs to act as network boosters, potentially reducing installation costs by up to 50\% while providing crucial reactive power compensation, high reliability, and low maintenance.
	
	Furthermore, beyond simply protecting the national grid from unbalanced loads, these advanced devices fundamentally revolutionize train operations. By synchronising their output for parallel feeding, SFC architectures facilitate the complete elimination of phase insulations, or "neutral sections", along the catenary without requiring extensive modifications to existing overhead line equipment (OLE) \cite{krastev_2016_future}. By deploying Active Power Flow Controllers (PFC) alongside the traction transformers, the grid-side power quality is actively managed, allowing the overhead line to provide continuous, uninterrupted power to the vehicles \cite{chen_2016_modelling}.

	Three architecturally distinct approaches to eliminating neutral sections should be distinguished, as they are occasionally conflated in the literature. The first is \textit{co-phase} traction power supply, which originates in Chinese research at Southwest Jiaotong University and operates at 25/27.5\,kV 50\,Hz. Co-phase schemes retain a conventional balance transformer but add a single-phase back-to-back converter that transfers active power between the feeding and compensating phases, simultaneously balancing active power, compensating reactive power and filtering harmonics, so that the substation presents a single phase to the catenary and the neutral section can be removed \cite{shu2011singlephase, shu2013digital, chen_2016_modelling}. The second is the centralised low-frequency railway grid used in the German-speaking and Nordic countries, which operates at 15\,kV 16.7\,Hz and is fed from a dedicated single-phase traction transmission network via rotary or static frequency converters; here the frequency conversion, rather than per-substation compensation, is the defining architectural feature. The third is SFC-fed 50\,Hz parallel feeding of the kind discussed above \cite{krastev_2016_future, bowdidge_2023_a}. These are distinct system families addressing overlapping problems, and the co-phase literature in particular should not be read as describing 16.7\,Hz practice. The algorithmic implications, however, converge: in all three cases the substation interface becomes a power-electronic device with hard current limits rather than a passive thermal one. The current state-of-the-art for these converters relies heavily on Modular Multilevel Converter (MMC) solutions. As reviewed by \cite{ronanki_2018_modular}, MMCs provide superior fault-tolerant capabilities; however, the highly complex internal capacitor voltage balancing makes them sensitive to sudden fluctuations. While these converter-based architectures represent the future of electrified rail, their reliance on semiconductor power electronics introduces a critical vulnerability: they possess absolute, hard current limits. If the instantaneous power demand of the trains within a sector exceeds these limits, the system will actively curtail power, reinforcing the need for intelligent, software-side load management.
	
	\begin{figure*}[!t]
	\centering
	\resizebox{0.95\textwidth}{!}{
		\begin{tikzpicture}[
			>=stealth, thick,
			gridline/.style={-, line width=1.5pt, color=gray},
			powerline/.style={->, line width=2pt},
			ole/.style={-, line width=2pt, color=blue!70!black},
			returnline/.style={-, line width=2pt, color=violet!75!black},
			raill/.style={-, line width=2pt, color=violet!75!black},
			block/.style={rectangle, draw, rounded corners, align=center, minimum height=1.5cm, font=\small},
			labelnode/.style={font=\bfseries\small, align=center},
			tag/.style={font=\scriptsize, fill=white, inner sep=1.5pt, align=center}
			]

			\node[labelnode] at (2.2, 5.5) {Traditional Transformer Architecture\\[1mm] \textit{(Unbalanced Load \& Neutral Sections)}};

			\draw[gridline, dashed, color=gray!45] (-1.2, 4.3) -- (5.5, 4.3) node[right, font=\footnotesize, color=gray!65] {L1};
			\draw[gridline] (-1.2, 3.9) -- (5.5, 3.9) node[right, font=\footnotesize] {L2};
			\draw[gridline] (-1.2, 3.5) -- (5.5, 3.5) node[right, font=\footnotesize] {L3};
			\node[font=\scriptsize\itshape, color=gray!65, anchor=east] at (5.3, 4.55) {not connected at this substation};

			\node[block, fill=red!10, minimum width=2.9cm, minimum height=1.4cm] (trafo) at (1.5, 1.8)
			{Traction Transformer\\[0.4mm] \textit{(single-phase, L2--L3)}};

			\draw[powerline, color=red!70!black] (1.0, 3.9) -- (1.0, 2.5);
			\draw[powerline, color=red!70!black] (2.0, 3.5) -- (2.0, 2.5);
			\node[tag] at (1.5, 3.02) {Unbalanced\\line-to-line draw};

			\draw[powerline, color=blue!70!black] (0.5, 1.1) -- (0.5, 0.06);
			\draw[returnline] (2.5, 1.1) -- (2.5, -1.3);
			\node[tag, anchor=west, text=violet!75!black] at (2.62, 0.92) {Secondary return};

			\draw[ole] (-1.2, 0) -- (3.6, 0);
			\node[tag, anchor=south, text=blue!70!black] at (1.5, 0.12) {$+25$\,kV OLE\\(Section A)};

			\draw[line width=4pt, color=orange] (3.6, 0) -- (4.2, 0);
			\node[font=\scriptsize\bfseries, color=orange!90!black, align=center, anchor=south, fill=white, inner sep=1.5pt] at (3.9, 0.12) {Neutral\\Section};

			\draw[ole, color=blue!35!black, dash pattern=on 5pt off 2.5pt] (4.2, 0) -- (5.5, 0);
			\node[tag, anchor=north] at (4.45, -0.14) {Section B: adjacent substation,\\rotated phase pair (L1--L2)};

			\draw[raill] (-1.2, -1.3) -- (5.5, -1.3);
			\node[tag, anchor=south west, text=violet!75!black] at (-1.15, -1.22) {Running rails (0\,V return path)};

			\draw[line width=1.2pt, color=violet!75!black] (2.5, -1.3) -- (2.5, -1.60);
			\draw[line width=1.2pt, color=violet!75!black] (2.18, -1.60) -- (2.82, -1.60);
			\draw[line width=1.2pt, color=violet!75!black] (2.30, -1.75) -- (2.70, -1.75);
			\draw[line width=1.2pt, color=violet!75!black] (2.42, -1.90) -- (2.58, -1.90);

			\node[labelnode] at (11.2, 5.5) {Static Frequency Converter (SFC)\\[1mm] \textit{(Balanced Load \& Continuous Catenary)}};

			\draw[gridline] (7.9, 4.3) -- (14.3, 4.3) node[right, font=\footnotesize] {L1};
			\draw[gridline] (7.9, 3.9) -- (14.3, 3.9) node[right, font=\footnotesize] {L2};
			\draw[gridline] (7.9, 3.5) -- (14.3, 3.5) node[right, font=\footnotesize] {L3};

			\node[block, fill=green!10, minimum width=3.9cm, minimum height=1.4cm] (sfc) at (11.2, 1.8)
			{SFC Power Electronics\\[0.4mm] \textit{(Rectifier $\rightarrow$ DC $\rightarrow$ Inverter)}};

			\draw[powerline, color=green!60!black] (10.2, 4.3) -- (10.2, 2.5);
			\draw[powerline, color=green!60!black] (11.2, 3.9) -- (11.2, 2.5);
			\draw[powerline, color=green!60!black] (12.2, 3.5) -- (12.2, 2.5);
			\node[tag] at (11.2, 3.02) {Balanced\\3-phase draw};

			\draw[powerline, color=blue!70!black] (10.4, 1.1) -- (10.4, 0.06);
			\draw[returnline] (13.0, 1.1) -- (13.0, -1.3);
			\node[tag, anchor=east, text=violet!75!black] at (12.88, 0.92) {Converter return};

			\draw[ole] (7.9, 0) -- (14.3, 0);
			\node[tag, anchor=south, text=blue!70!black] at (11.7, 0.12) {Continuous $+25$\,kV catenary\\(no phase insulations)};

			\draw[raill] (7.9, -1.3) -- (14.3, -1.3);
			\node[tag, anchor=south west, text=violet!75!black] at (7.95, -1.22) {Running rails (0\,V return path)};

			\draw[line width=1.2pt, color=violet!75!black] (13.0, -1.3) -- (13.0, -1.60);
			\draw[line width=1.2pt, color=violet!75!black] (12.68, -1.60) -- (13.32, -1.60);
			\draw[line width=1.2pt, color=violet!75!black] (12.80, -1.75) -- (13.20, -1.75);
			\draw[line width=1.2pt, color=violet!75!black] (12.92, -1.90) -- (13.08, -1.90);

			\draw[dashed, color=gray, line width=1pt] (6.85, 6.1) -- (6.85, -2.1);

		\end{tikzpicture}
	}
	\caption{Comparative schematic of traction grid integration. In the conventional arrangement (left) the transformer primary is connected line-to-line across two of the three transmission phases, the third being unused at that substation; the secondary supplies the catenary at one terminal and is bonded to the running rails and earthing network at the other, so the rail connection is a property of the secondary return path rather than a connection of the unused third phase. Adjacent substations are supplied from rotated phase pairs, distributing the unbalance across all three phases in aggregate, and the resulting phase difference between neighbouring sections necessitates the neutral section shown. Static Frequency Converters (right) draw a balanced three-phase load and permit continuous parallel feeding to the heavy-duty fleet without phase insulations. Blue denotes the catenary supply path and violet the return path through the running rails. Balanced special connections (V/v, Scott, YNd11, YNvd, impedance-matching), which do use all three grid phases, are discussed in the text.}
	\label{fig:sfc_comparison}
\end{figure*}
	
	\subsection{Regenerative Braking and Energy Storage Systems (ESS)}
	
	Alongside SFCs, the integration of off-board Energy Storage Systems (ESS) and local renewable energy sources presents a significant hardware mitigation strategy for grid capacity limits. Off-board ESS can capture surplus energy generated during a train's regenerative braking phase or from local renewables, storing it to assist locomotive acceleration during peak demand or to supply railway stations. By localising the energy source, ESS deployments can significantly reduce the instantaneous overload on the national grid, theoretically enabling the integration of more or larger electric locomotives into the existing network.
	
	However, the operational and economic viability of ESS is highly dependent on traffic density. As demonstrated by operational data from the East Japan Railway Company (Hitachi Branch) \cite{barros_2020_opportunities}, in metropolitan areas with high-density traffic (headways of under 5 minutes), the energy from regenerative braking is typically consumed instantaneously by adjacent accelerating locomotives, rendering stationary ESS redundant. Conversely, in rural areas where train headways exceed 10 minutes, the infrequency of regenerative events often fails to justify the capital investment of ESS infrastructure.
	
	Consequently, the primary challenge in exploiting regenerative braking lies not just in the storage hardware, but in the sophisticated control algorithms required to detect braking events across tens of kilometres and intelligently route that power. This limitation reinforces the necessity for network-wide Energy Management Strategies (EMS) capable of dynamically synchronising train trajectories, thereby maximising the natural, real-time energy exchange between vehicles before relying on expensive stationary storage.
	
	\subsection{Onboard Energy Buffers as a Mitigation Strategy}
	
	Beyond stationary grid upgrades such as SFCs and trackside ESS, another hardware mitigation strategy involves onboard energy buffers. Battery-electric and hydrogen-bi-mode rolling stock has been proposed as a way to provide operational independence from the catenary during high-demand events, by allowing a vehicle entering a saturated electrical sector to temporarily reduce or eliminate its grid draw. The detailed thermochemical and lifecycle properties of these onboard buffers fall outside the algorithmic-power-optimisation scope of this review and are addressed at length elsewhere. From an algorithmic perspective, the key implication is that the additional onboard degree of freedom expands the EMS decision space without removing the underlying grid constraint. This is reinforced normatively: clause 7.1 of \cite{en503881} includes 
	charging current or power to on-board energy storage systems within the traction current or power exchanged through the pantograph, and clause 7.2 defines the maximum current as the total of traction, auxiliary and energy storage demand. Recharging a buffer therefore consumes the same constrained pantograph current that traction does, so the buffer relocates demand in time rather than removing it from the feeding section, and consequently amplifies rather than reduces the need for the dynamic, multi-objective control approaches reviewed in subsequent sections.
	
	\section{Power Quality and Network Capacity Challenges}
	\label{sec:power_quality}
	Despite the integration of the mitigation technologies reviewed in Section \ref{electrification_systems} such as auto-transformer feeding arrangements, converter-based substations including static frequency converters and co-phase architectures, off-board energy storage, and onboard energy buffers, the fundamental electrical characteristics of heavy-duty railway traffic continue to pose severe power quality challenges. While electrification studies often treated capacity as a static metric, recent reviews by \cite{hayashiya_2020_recent}, \cite{barros_2020_opportunities}, and \cite{he_2016_power} identify a fundamental paradigm shift: traction loads are uniquely difficult to manage because they are highly dynamic, massive in scale, and fluctuate dramatically based on instantaneous vehicle acceleration. Building on this, \cite{serranojimnez_2017_electrical} critiques traditional transformer-based architectures, arguing they have reached a hard ceiling regarding the amount of power they can deliver, which directly causes significant power quality degradation.
	
	\subsection{Harmonics, Phase Unbalance, and Hardware Mitigation limits}
	
	The widespread adoption of modern power electronics has created a paradox within the literature. Vehicle-centric studies credit traction inverters with substantial improvements in traction-system efficiency, and the mechanism is worth stating because it is the same mechanism that creates the difficulty on the network side. Variable-voltage variable-frequency inverter drives allow an AC traction motor to be operated close to its peak-efficiency across the whole speed range, rather than dissipating surplus energy in resistive control as earlier rheostatic schemes did, and their four-quadrant 
	capability is what allows regenerative braking to return energy to the catenary at all. 
	Both advantages depend on high-frequency pulse-width modulation of the converter, and it 
	is precisely this switching action that infrastructural researchers such as 
	\cite{he_2016_power} identify as degrading the grid, injecting high-frequency harmonics 
	back into the overhead line. The efficiency gain and the power quality penalty therefore 
	come from the same component, which is why the tension cannot be resolved at vehicle 
	level alone. Furthermore, \cite{tanta2018power} highlights that the interconnection of a highly fluctuating single-phase traction system to the three-phase Public Power System (PPS) inherently generates Negative Sequence Components (NSCs).
	
	To address this, authors such as \cite{anluo_2011_railway} advocate for dedicated hardware compensators, such as Railway Static Power Conditioners (RPCs). However, the literature notes that RPCs and SFCs rely on back-to-back converters with finite thermal and current capacities. If multiple trains accelerate simultaneously within the same feeding zone, the resulting instantaneous power draw easily exceeds these limits, exposing the fragility of relying solely on localised substation hardware.
	
	\subsection{Low Voltage Behaviour and Infrastructure Tripping}
	
	Beyond harmonics, the literature identifies physical voltage drop as the most immediate operational constraint restricting network capacity. While earlier studies often treated voltage sag as a passive network characteristic resulting from heavy current sinks, work by Tian \cite{tian2017system} shifted the paradigm by mathematically demonstrating that the grid and the vehicle are dynamically coupled. The train--network coupling principle established there was developed for DC traction networks; its transfer to the AC 25\,kV case considered in this review is supported by AC-specific system studies including \cite{pan2020integrated} and \cite{tao_2024_comprehensive}, and by the normative voltage-dependent current limitation of \cite{en503881}, which applies to AC traction irrespective of the modelling approach.
	
	Rather than viewing the train as an independent actor, Tian demonstrates that a train's maximum mechanical traction power is strictly dependent on the instantaneous pantograph voltage. This dependency is not a design choice of individual manufacturers but a normative requirement. Clause 7.3 of BS EN 50388-1 \cite{en503881} requires train sets to be equipped with an automatic function that adapts maximum steady-state power or current consumption to line voltage, with the governing factor $a$ fixed at 0.90 for AC 25\,kV 50\,Hz systems in Table 2 of that clause. The requirement carries no lower power threshold and therefore applies across the whole fleet: the 2\,MW exemption in clause 7.2 concerns a distinct function, namely limitation of the current exchange to a fixed value supplied by the infrastructure manager through the register of infrastructure, the route description or ETCS, and does not exempt any vehicle from the voltage-dependent characteristic. Traction power at wheel and traction current are required to be zero at $U_{min2}$, and the limitation may be realised on either the current or the power. Where current limitation is used, maximum traction current is derived from the maximum power at wheel at $U_n$ and is limited linearly at voltages 
	below $a\,U_n = 22.5$\,kV. Where power limitation is used, the maximum power at wheel is limited approximately linearly in two steps: from $P_{max}$ at $U_n$ to $a\,P_{max}$ at $a\,U_n$, and from $a\,P_{max}$ to zero at $U_{min2} = 17.5$\,kV. Both forms are presented graphically in normative Annex F of \cite{en503881}. The characteristic governs traction only: it is expressly inapplicable in regenerative 
	braking mode and does not affect auxiliary power consumption, although auxiliary load may be shed below $U_{min2}$.

	This characteristic has a natural per-unit reading, since the factor $a$ is dimensionless: limitation begins at $a = 0.90$\,p.u. ($22.5$\,kV) and traction reaches 
	zero at $U_{min2} = 0.70$\,p.u. ($17.5$\,kV) on a 25\,kV base. The values of 0.9\,p.u. and 0.7\,p.u. commonly encountered in practice therefore correspond directly to the 
	normative characteristic of \cite{en503881} rather than representing an informal convention. The lowest permanent voltage $U_{min1} = 19$\,kV specified by 
	\cite{en50163}, approximately 0.76\,p.u., lies between these two points, but it is a network-side compliance limit rather than a breakpoint of the rolling-stock 
	characteristic; a train may therefore be operating under current or power limitation, with substantially reduced tractive effort, while the network is still nominally within 
	its permitted range. Fig. \ref{fig:voltage_drop} is plotted against the standardised absolute thresholds with a secondary ordinate giving their per-unit equivalents, so 
	that both conventions can be read from the same figure. The literature links this hardware self-preservation directly to sluggish acceleration and missed timetable targets.
	
	Furthermore, Tian's models show that when multiple heavy trains attempt to accelerate simultaneously, they actively starve each other of power, frequently triggering substation circuit breakers. Consequently, authors such as \cite{barros_2020_opportunities} conclude that this low voltage behaviour acts as a hard ceiling on network capacity, observing that integrating new heavy-freight locomotives cannot rely on hardware upgrades alone.
	
	\begin{figure}[H]
	\centering
	\begin{tikzpicture}
		\begin{axis}[
			scale only axis,
			width=0.70\columnwidth,
			height=5.6cm,
			axis y line*=left,
			axis x line*=bottom,
			xlabel={Distance from Substation (km)},
			ylabel={Catenary Voltage (kV)},
			xmin=0, xmax=20,
			ymin=16, ymax=27,
			grid=major,
			legend style={at={(0.5,-0.26)}, anchor=north, legend columns=2, draw=none,
				column sep=1.2ex, font=\scriptsize},
			label style={font=\small},
			tick label style={font=\footnotesize}
			]

			\addplot[dashed, thick, black, domain=0:20] {25};
			\addlegendentry{$U_n$: 25\,kV (1.00\,p.u.)}

			\addplot[dashed, thick, red, domain=0:20] {22.5};
			\addlegendentry{$a\,U_n$: 22.5\,kV (0.90\,p.u.)}

			\addplot[dashdotted, thick, olive, domain=0:20] {19};
			\addlegendentry{$U_{min1}$: 19\,kV (0.76\,p.u.)}

			\addplot[dotted, very thick, orange!85!black, domain=0:20] {17.5};
			\addlegendentry{$U_{min2}$: 17.5\,kV (0.70\,p.u.)}

			\addplot[smooth, thick, blue] coordinates {
				(0, 25.0) (3, 24.6) (6, 23.9) (9, 22.8) (10.5, 20.5)
				(12, 17.5) (13.5, 19.8) (16, 21.0) (20, 20.5)
			};
			\addlegendentry{Pantograph voltage}

			\node[coordinate] (dip) at (axis cs:12, 17.5) {};
			\node[coordinate] (textpos) at (axis cs:6.6, 20.2) {};
			\draw[->, thick, shorten >=3pt] (textpos) -- (dip);
			\node[above, align=center, font=\scriptsize, fill=white, inner sep=2pt]
			at (textpos) {Heavy Freight\\Acceleration Event};
		\end{axis}

		\begin{axis}[
			scale only axis,
			width=0.70\columnwidth,
			height=5.6cm,
			axis y line*=right,
			axis x line=none,
			xmin=0, xmax=20,
			ymin=0.64, ymax=1.08,
			ylabel={Catenary Voltage (p.u. on 25\,kV base)},
			ytick={0.70, 0.76, 0.90, 1.00},
			yticklabels={0.70, 0.76, 0.90, 1.00},
			label style={font=\small},
			tick label style={font=\footnotesize}
			]
		\end{axis}
	\end{tikzpicture}
	\caption{Simulated Train-Track-Power (TTP) voltage profile, shown against both the absolute standardised thresholds (left ordinate) and their per-unit equivalents on a 25\,kV base (right ordinate). A sudden, massive current draw from an accelerating heavy freight train depresses the localised catenary voltage below $a\,U_n = 22.5$\,kV (0.90\,p.u.), at which point the automatic current or power limitation of BS EN 50388-1 \cite{en503881} clause 7.3 begins to reduce the train's permissible draw, and subsequently to $U_{min2} = 17.5$\,kV (0.70\,p.u.), at which traction power and current are required to be zero. The lowest permanent voltage $U_{min1} = 19$\,kV of BS EN 50163 \cite{en50163} is shown for reference as a network-side compliance limit. The profile shown is the unmitigated demand-side response absent coordinated energy management; where automatic limitation is active the sag is arrested by the train's own reduction in draw, at a corresponding cost in tractive effort and schedule adherence.}
	\label{fig:voltage_drop}
\end{figure}
	
	\subsection{Grid Balancing and Reactive Power Compensation}
	
	To counteract these severe low-voltage behaviours, the literature shows a clear historical progression in grid balancing strategies. Because heavy traction units act as inductive loads, they draw substantial magnetising currents that degrade the localized power factor and pull down catenary voltage.
	
	Historically, the primary hardware mitigation for this imbalance has been the installation of trackside capacitive compensation equipment. A prominent real-world application is found on the UK's High Speed 1 (HS1) route, where inherent transformer design caused the nominal 25kV supply to sag to critical 17.5kV thresholds \cite{a2013_abb}. To restore grid balance, capacitive filters were deployed at strategic substations to artificially inject reactive power, stabilizing the catenary voltage.
	
	As power quality constraints have tightened, this trackside hardware has evolved into more complex active systems. For instance, \cite{maghsoud2013current} demonstrate the deployment of Traction Power Conditioners (TPCs) coupled with specialized $\Upsilon/\Delta$ transformers to actively shift active and reactive power between adjacent electrical sections, simultaneously mitigating both reactive power deficits and Negative Sequence Currents (NSC).
	
	However, while trackside Static Var Compensators (SVCs) and TPCs are effective at localised grid balancing, they represent spatially fixed, capital-intensive solutions. The literature shows a contemporary research shift away from trackside hardware toward distributed, train-level software coordination. A 2020 study by \cite{morais2020new} explicitly compared the deployment of trackside SVCs against dynamic train-level reactive power adaptation, finding that by commanding the trains' onboard converters to dynamically adjust their own reactive power output, the theoretical capacity of the railway infrastructure could be increased by 50\%—matching the performance of a multi-million-pound trackside SVC with less than a 10\% increase in the train's apparent power.
	
	Two distinct senses of ``distributed'' are in use here and are worth separating. In 
	\cite{morais2020new} the compensating devices themselves are distributed: the reactive 
	power is supplied by the onboard converters of the trains, which are already present 
	throughout the network, and the control decision is taken locally by each vehicle. A 
	second architecture distributes the devices but centralises the decision. As demonstrated 
	by \cite{hao2021distributed}, a central dispatch algorithm can allocate reactive power 
	set-points across substations that are physically spread along the route. What both share, 
	and what distinguishes them from a trackside SVC, is that the compensating capability is 
	distributed in space and commanded in software rather than concentrated in a single fixed 
	installation.
	
	This is also what allows existing assets to be repurposed rather than supplemented. 
	Reversible substations are installed to return regenerative braking energy to the upstream 
	supply, and their converters are rated for that duty; outside recovery events, a proportion 
	of that converter rating is idle. Because a four-quadrant converter can exchange reactive 
	power independently of the active power it is transferring, that idle apparent-power 
	capacity can be commanded to provide compensation without additional hardware, subject 
	only to the converter's current and thermal limits. The compensation is therefore obtained 
	from assets already justified on other grounds, which is the economic argument the 
	literature makes for the approach. The scheme in \cite{hao2021distributed} is developed 
	for DC traction supply; the converter capability it exploits is not specific to the supply 
	type, but the quantitative results reported there are not transferred to the AC case here.
	
	The literature thus indicates that as uncoordinated heavy-freight acceleration events become more frequent, over-dimensioning static capacitive hardware to handle every transient peak becomes financially inefficient relative to coordinated software-side control. Several recent studies suggest that multi-train EMS approaches can extract additional capacity from existing grid hardware by flattening the peak load curve, although the comparative effectiveness of such approaches remains an active area of investigation.
	
	\section{The Mixed Fleet Traffic Problem: Freight vs. Passenger Integration}
	\label{sec:mixed_fleet}
	
	While hardware limitations and generalised power quality issues affect all electrified routes, the most severe operational bottlenecks occur on mixed-traffic networks. Unlike dedicated high-speed lines, networks such as the UK mainline must support highly mixed fleet traffic, integrating lightweight, fast-accelerating passenger multiple units with massive, slow-accelerating freight locomotives. The contrasting electrical demand profiles produced by these two vehicle classes are shown in Fig. \ref{fig:mixed_traffic_load}, and the resulting multi-train regenerative power flow is illustrated in Fig. \ref{fig:regen_power_flow}.
	
	\begin{figure}[H]
	\centering
	\begin{tikzpicture}
		\begin{axis}[
			width=0.96\columnwidth,
			height=5.5cm,
			xlabel={Time (Seconds)},
			ylabel={Instantaneous Power (MW)},
			xmin=0, xmax=180,
			ymin=0, ymax=10,
			grid=major,
			legend style={at={(0.5,-0.24)}, anchor=north, legend columns=2, draw=none,
				column sep=1.8ex, font=\scriptsize},
			label style={font=\small},
			tick label style={font=\footnotesize}
			]

			\addplot[smooth, thick, blue] coordinates {
				(0, 0)
				(15, 8.5)
				(30, 2.0)
				(90, 2.0)
				(180, 2.0)
			};
			\addlegendentry{Passenger Multiple Unit}

			\addplot[smooth, thick, red, dashed] coordinates {
				(0, 0)
				(30, 6.0)
				(60, 7.5)
				(120, 7.0)
				(160, 5.0)
				(180, 4.0)
			};
			\addlegendentry{Heavy Freight Locomotive}

			\node[align=center, font=\scriptsize, text=blue, fill=white, inner sep=1pt] at (axis cs:40, 8.8) {Transient Spike};
			\node[align=center, font=\scriptsize, text=red, fill=white, inner sep=1pt] at (axis cs:110, 8.8) {Continuous Sustained Load};

		\end{axis}
	\end{tikzpicture}
	\caption{Contrasting electrical demand profiles in a mixed-traffic network. While lightweight passenger units rely on highly transient power spikes, heavy freight locomotives mandate continuous, massive current draws over prolonged time windows to overcome inertia and gradient resistance.}
	\label{fig:mixed_traffic_load}
\end{figure}
	
	\begin{figure*}[!t]
		\centering
		
		\resizebox{0.8\textwidth}{!}{
			\begin{tikzpicture}[
				>=stealth, 
				trainbox/.style={rectangle, draw, rounded corners, minimum height=1.5cm, align=center, font=\small},
				powerarrow/.style={->, line width=2.5pt},
				regenarrow/.style={->, line width=2.5pt, color=green!60!black},
				gridarrow/.style={->, line width=2.5pt, color=red!70!black}
				]

				\draw[ultra thick] (0, 4) -- (12, 4) node[right, font=\bfseries] {+25kV Catenary (OLE)};
				
				\draw[ultra thick] (0, 0) -- (12, 0) node[right, font=\bfseries] {0V Running Rails};

				\node[draw, rectangle, fill=blue!10, minimum height=4cm, minimum width=1.8cm, align=center, font=\bfseries] (sub) at (0, 2) {TPSS\\(Grid)};

				\node[trainbox, minimum width=3.5cm, fill=red!5] (freight) at (4.5, 0.8) {\textbf{Heavy Freight}\\(Accelerating)};

				\node[trainbox, minimum width=3cm, fill=green!5] (passenger) at (9.5, 0.8) {\textbf{Passenger Unit}\\(Braking)};

				\draw[thick] (freight.north) -- (4.2, 2.5) -- (4.8, 3.2) -- (4.5, 4);
				\draw[thick] (passenger.north) -- (9.2, 2.5) -- (9.8, 3.2) -- (9.5, 4);

				\draw[gridarrow] (1.2, 4.3) -- node[above, font=\small, text=red!70!black] {Grid Power ($P_{grid}$)} (4, 4.3);

				\draw[regenarrow] (10.3, 1.8) -- node[right, font=\small, text=green!60!black, align=center] {Regenerative\\Power ($P_{regen}$)} (10.3, 3.8);

				\draw[regenarrow] (8.8, 4.3) -- node[above, font=\small, text=green!60!black] {Recuperated Power} (5, 4.3);

				\draw[powerarrow, color=orange!80!black] (3.8, 3.8) -- node[left, font=\small, text=orange!80!black, align=center] {Massive\\Traction\\Demand} (3.8, 1.8);

				\node[align=center, font=\small] at (4.5, -0.6) {$\mathbf{P_{demand} = P_{grid} + P_{regen}}$};
				
			\end{tikzpicture}
		}
		\caption{Multi-train regenerative power flow. In a grid-aware Train-Track-Power (TTP) simulation, the kinetic energy recuperated by a braking passenger unit is actively consumed by an accelerating heavy freight locomotive within the same electrical sector, dynamically reducing the total peak power drawn from the substation.}
		\label{fig:regen_power_flow}
	\end{figure*}
	
	\subsection{The Continuous Load Profile of Heavy Freight}
	
	The literature frequently highlights the incompatibility of these two electrical demand profiles when unmanaged. A passenger train requires a brief, sharp spike in current to reach its cruising speed, after which its power demand drops considerably. In contrast, a fully laden freight train (often exceeding 2,000 tonnes) requires a massive, sustained current draw over a prolonged time window just to overcome its initial inertia and gradient resistance.
	
	A critical oversight in traditional infrastructure design, as identified by \cite{femine_2020_power}, is the reliance on averaged public grid loads, which fail to adequately account for the highly transient, sustained traction demands of heavy freight. Building on this critique, \cite{hu_2017_a} demonstrates that the electrical burden extends beyond raw current draw: mixed-freight fleets frequently operate at severely degraded power factors (often between 0.70 and 0.84), amplifying the injection of negative sequence components and reactive power back into the national grid.
	
	Recent research therefore frames the accelerating heavy freight train not only as a vehicle, but as a continuous load sink that depresses the localised catenary voltage for miles around it.
	
	\subsection{Simultaneous Traction Overloads and Capacity Starvation}
	
	This continuous voltage drop creates a severe operational hazard for other vehicles sharing the same electrical section. If a passenger train enters the section and attempts to accelerate while the freight train is still drawing peak current, the combined load forces the line voltage below the critical 19kV threshold.
	
	While early timetabling algorithms treated train movements purely as a spatial and temporal problem, recent Train-Track-Power (TTP) models challenge this approach. As modelled by \cite{tao_2024_comprehensive}, train operation that ignores the TTP grid interaction directly causes continuous voltage fluctuations that starve trailing trains of power and trip substation circuit breakers. The mechanism by which coordination reduces that burden is summarised in Fig. 
	\ref{fig:regen_power_flow}. When a braking vehicle and an accelerating vehicle occupy the 
	same electrical section simultaneously, the recuperated power $P_{regen}$ is consumed 
	locally by the accelerating train, and the substation supplies only the shortfall, 
	$P_{grid} = P_{demand} - P_{regen}$. The peak drawn from the substation is therefore set 
	not by the largest instantaneous traction demand in the section but by the largest 
	\emph{uncompensated} demand. What multi-train coordination contributes is the coincidence 
	itself: in uncoordinated operation a braking event and an accelerating event align in time 
	and electrical section only incidentally, and where they do not, the recuperated energy 
	finds no receptive load and is dissipated in on-board or line resistors. By scheduling 
	acceleration phases against the braking phases of other services in the same section, an 
	EMS converts an incidental overlap into a designed one, and the reduction in peak 
	substation draw follows directly.
	
	Tao's work effectively redefines the concept of network capacity. By illustrating this phenomenon of simultaneous traction overloads, the literature shows that physical track capacity (signalling block distance) is no longer the primary constraint on railway timetabling for many mixed-traffic networks; electrical capacity is. Because the infrastructural literature has established that upgrading the physical copper and transformers to handle these combined loads is economically prohibitive, authors such as \cite{tao_2024_comprehensive} position dynamic load shifting through software-side control as a primary research focus for managing voltage stability under mixed-traffic conditions.
	
	\section{Algorithm Family I: Analytical and Exact Methods} \label{sec:limitations}
	
	This section reviews the first algorithm family identified by the classification framework introduced in Section \ref{sec:taxonomy}: analytical methods and exact discrete methods. These approaches share a common property: they aim to derive a mathematically guaranteed global optimum, at the cost of imposing strong structural assumptions on the problem. Representative studies, their reported scope, and the limitations identified by their authors are summarised in Table \ref{tab:algorithm_evolution}.
	
	\subsection{The Rigidity of Analytical and Convex Optimization Models}
	
	Optimisation of train trajectories and energy consumption has historically relied on deterministic mathematical frameworks, primarily utilising techniques such as the Calculus of Variations, Pontryagin's Maximum Principle (PMP), and Mixed-Integer Linear Programming (MILP). These analytical models are highly effective at finding mathematically exact global minima for energy use, provided the operational environment is highly controlled and static.
	
	By comparing these methods, the literature reveals distinct operational contexts for each. MILP excels at high-level, discrete timetable scheduling across a network, whereas PMP is widely used for deriving exact continuous vehicle control equations. For example, foundational studies such as \cite{optimal_control_of_train} demonstrate how establishing "necessary conditions" via mathematical proofs can dictate the most efficient driving strategy for a given route.
	
	However, the primary limitation of these approaches in modern multi-train, grid-coupled 
	settings is structural rather than procedural. A distinction should be drawn within the 
	family: analytical methods such as PMP and the calculus of variations do yield closed-form 
	necessary conditions, whereas MILP and convex programming are solved by iterative numerical 
	procedures, simplex or interior-point iterations, and branch-and-bound search over 
	integer variables. The rigidity at issue is therefore not that the solvers are 
	non-iterative, but that each method admits only problems already expressed in a particular 
	structural form. PMP requires a differentiable Hamiltonian and well-posed boundary 
	conditions; MILP requires the model to be linear in its continuous variables; convex 
	programming requires a convex objective over a convex feasible set. Introducing 
	relationships that violate the required form does not slow these methods down so much as 
	place the problem outside the class they are guaranteed to solve. As noted by \cite{local_energy_minimisation}, these models rely on strict convexity assumptions: convex optimisation requires the objective function and constraints to form a convex set, meaning the relationships between variables must remain relatively predictable and continuous.
	
	When the physical system behaves unpredictably, the underlying mathematics break down. Integrating dynamic constraints such as real-time grid power headroom, fluctuating substation thermal limits, or stochastic mixed-traffic delays frequently challenges the convexity of the problem. Three distinct responses to this difficulty are conflated in parts of the literature and are worth separating, since they have different consequences for tractability. \emph{Linearisation} replaces a nonlinear relation with an affine one, preserving linear programming structure at the cost of an approximation error that is bounded only locally. \emph{Piecewise-affine approximation} improves that accuracy by using several segments, but it preserves convexity only where the approximated function is itself convex; 
	representing a non-convex function exactly in this way requires binary or SOS2 variables, 
	which moves the problem into the mixed-integer class and replaces polynomial-time solution 
	with branch-and-bound search. \emph{Convexification} is different in kind: it reformulates 
	or relaxes the problem, by change of variables, by McCormick envelopes, or by 
	second-order cone or semidefinite relaxation so that the feasible set is genuinely 
	convex, which restores a global optimality guarantee for the relaxed problem but not 
	necessarily a tight bound on the original one. Piecewise linearisation, in other words, 
	does not by itself make a non-convex problem convex.
	
	Railway formulations use all three. In \cite{mathematical_modeling_for_holistic_convex_optimisation}, the vehicle physics are linearised so that the resulting programme falls within the convex class the chosen solver requires. Modern MILP and convex solvers handle such approximations well, and topography 
	in particular can be represented with considerable precision by this route. The literature 
	increasingly critiques the approach, however, when it is applied to highly stochastic, 
	multi-agent grid-coupled environments. Linearising the physics strips away the dynamic real-world complexities, such as sudden grid power curtailments and the non-linear response of trains to voltage sags, that strictly dictate actual energy consumption in mixed-fleet networks.
	
	\subsection{Establishing the Theoretical Benchmark: The Optimal Control Sequence}
	
	Despite their lack of real-time adaptability and reliance on simplified physics, analytical models remain valuable in the literature for establishing theoretical baselines. They establish the minimum-energy trajectory under a specific and restrictive set of assumed conditions: a single train considered in isolation; a deterministic and fully known gradient profile and speed-limit sequence; a fixed journey time; constant or known vehicle mass; and an ideal power supply that imposes no voltage or current 
	constraint on the achievable tractive effort. Under those conditions the resulting 
	trajectory is a genuine optimum, and it is valuable precisely as a reference: the 
	assumptions it relies on are the ones that the multi-train and grid-coupled formulations 
	reviewed in Sections \ref{sec:heuristics} and \ref{sec:predictive} are obliged to relax.
	
	A significant 2013 study, \cite{energy_efficient_train_control}, successfully factored in static constraints like steep gradients and fixed speed limits to mathematically demonstrate a fundamental operational principle: the most energy-efficient trajectory for a rail vehicle generally follows a specific four-phase sequence: \textit{Maximum Power (Acceleration), Speed-Hold (Cruising), Coasting, and Maximum Braking}.
	
	This sequence serves as a benchmark for subsequent optimisation research. While analytical calculus requires a highly precise, inflexible formulation of the track that is difficult to adapt to dynamic variables, the "Power-Hold-Coast-Brake" profile itself is widely accepted as the standard of efficient driving behaviour. However, recent TTP studies argue that while this sequence is optimal for a single train in isolation, it can fail in multi-train networks where dynamic grid constraints force vehicles to brake or coast prematurely. Nonetheless, when a topography-aware meta-heuristic produces a multi-objective trajectory that organically mirrors this sequence without being hard-coded to do so, its underlying driving strategy is consistent with the theoretical benchmark.
	
	\subsection{Dynamic Programming and the "Curse of Dimensionality"}
	
	To bridge the gap between pure analytical calculus and complex discretised track constraints, much of the literature has turned to Dynamic Programming (DP). Building on Bellman's Principle of Optimality, DP discretises the journey into finite distance or time steps, creating a grid of possible states (e.g., speed, position, and time). By working backward from the destination to the origin, DP evaluates every possible state transition to guarantee the absolute global optimum for energy consumption.
	
	When compared to convex optimisation, DP is widely cited for its ability to handle non-linear physics without aggressive simplification. For example, the 2013 paper \cite{single_train_trajectory_optimisation} utilises DP to successfully navigate discrete distance steps and speed profiles, demonstrating its ability to handle precise topographical variations that continuous analytical models struggle with. Similarly, studies such as \cite{a_dynamic_programming_approach} highlight DP's ability to map out exact speed constraints.
	
	However, the literature consistently identifies a flaw when applying DP to modern grid-coupled multi-train networks: the exponential increase in computational cost, universally referred to as the "curse of dimensionality". DP algorithms must evaluate every possible combination of states in the grid. In a single-train, mechanical-constraint-only model, the state variables are relatively limited (speed, distance, mass, gradient). In a multi-train, grid-coupled model, the state space expands drastically: the model must simultaneously represent multiple train positions, multiple train speeds, the catenary voltage in each electrical sector, instantaneous substation loading, and the resulting dynamic grid power headroom available to each vehicle.
	
	As the 2013 \cite{single_train_trajectory_optimisation} study explicitly notes, the computational cost and calculation time increase dramatically as more state variables and constraints are added. The grid becomes too dense to compute efficiently. While DP is mathematically rigorous and provides a strong offline baseline, it frequently requires hours to compute a single journey for a complex multi-train scenario, rendering it unsuited for the real-time, dynamic decision-making that operational grid management requires.
	
	\subsection{The Transition to Predictive and Real-Time Systems}
	
	Because analytical models impose restrictive structural assumptions and Dynamic Programming carries prohibitive computational cost, the literature shows a clear progression toward alternative Energy Management Strategies. Early attempts to satisfy the real-time requirement included methods such as the Equivalent Consumption Minimisation Strategy (ECMS) and a partial return to rule-based logic.
	
	However, the literature documents a clear divide. While ECMS has proven successful in the automotive sector for lightweight passenger vehicles on relatively flat roads, it has been found to perform less well in heavy-duty rail because it lacks rigorous topography-awareness. Rule-based strategies more generally are inherently static: they react to the vehicle's current state but cannot predict upcoming topography or anticipate grid constraints.
	
	To overcome the curse of dimensionality without linearising the physics, the literature points toward heuristic and meta-heuristic approaches, which are reviewed in Section \ref{sec:heuristics}. These methods reduce computational cost relative to exhaustive DP searches, but the literature notes that evaluating massive populations of Train-Track-Power (TTP) simulations still requires minutes or hours, positioning meta-heuristics as offline or day-ahead planners rather than real-time controllers.
	
	\section{Algorithm Family II: Meta-Heuristic Methods}
	\label{sec:heuristics}
	
	\subsection{Overcoming the Computational Limitations}
	
	As established in Section \ref{sec:limitations}, while deterministic models and Dynamic Programming (DP) offer mathematical certainty, their exponential computational cost limits their applicability to live, dynamic networks. A comprehensive 2017 survey, \cite{review_of_energy_efficient_train_control}, highlights this core trade-off in the literature: the tension between model accuracy (complex physics) and computation speed (real-time application). The review notes that highly theoretical models frequently fail in practice because they are too slow for on-board hardware to process.
	
	The analytical and exact methods reviewed in Section \ref{sec:limitations} deliver 
	optimality guarantees at a computational cost that scales poorly with fleet size and 
	constraint coupling, placing them outside the window required for operational deployment. 
	To bridge this gap between theoretical physics and real-time operational capability, the 
	literature points toward heuristic and meta-heuristic solvers. Unlike deterministic 
	algorithms that evaluate every state transition in a rigid grid, heuristics utilise 
	stochastic search processes to explore the solution space, mitigating the ``curse of 
	dimensionality'' that restricts DP models. The meta-heuristic studies reviewed in this 
	section, together with the analytical and exact methods discussed in Section 
	\ref{sec:limitations}, are catalogued in Table \ref{tab:algorithm_evolution}.

	\begin{table*}[!t]
		\centering
		\caption{Comparative Analysis of Optimisation Algorithms in Grid-Constrained Railway Networks}
		\label{tab:algorithm_evolution}
		\small 
		\begin{tabularx}{\textwidth}{@{} 
				p{3.5cm} 
				p{2.5cm} 
				p{2.5cm} 
				X        
				X @{}}   
			
			\toprule
			\textbf{Study (Year)} & \textbf{Scope} & \textbf{Optimisation Method} & \textbf{Key Constraints Managed} & \textbf{Identified Limitations} \\
			\midrule

			\textit{The Optimal Control of a Train} (2000) \cite{optimal_control_of_train} & Single Train & Pontryagin's Maximum Principle (PMP) & Fuel minimisation, variable gradients, journey time. & Derives rigid "necessary conditions". Hard to adapt to complex, non-linear variables.\\
			\addlinespace
			
			\textit{Energy-efficient train control} (2013) \cite{energy_efficient_train_control} & Passenger \& Freight Rail & Calculus of Variations & Steep gradients (topography), speed limits. & Requires highly precise mathematical formulation of the track; lacks real-time adaptability.\\
			\addlinespace
			
			\textit{Mathematical Modeling for Holistic Convex...} \cite{mathematical_modeling_for_holistic_convex_optimisation} & Hybrid Trains & Convex Optimisation & Dynamic power split, basic energy limits. & Required heavy linearisation of vehicle physics to ensure optimisation could function.\\
			\addlinespace
			
			\textit{A dynamic programming approach...} (2017) \cite{a_dynamic_programming_approach} & Single Train & Dynamic Programming (DP) & Variable gradients, speed limits, safety data. & Shows limitations on steep gradients where steady-state cruising is physically impossible.\\
			\addlinespace
			
			\textit{Convex Optimisation of Speed and Energy...} \cite{convex_optimisation_ems} & Fuel Cell Hybrid Train & Convex Optimisation & Speed limits, energy management. & Explicitly admits the model lacks dynamic "battery temperature constraints".\\
			\midrule

			\textit{Coast Control of Train Movement...} (2004) \cite{coast_control_with_ga} & Mass Transit (ATO) & Hierarchical Genetic Algorithm (HGA) & Schedule regulation, ride comfort via coasting. & Complex multi-point coasting too difficult for human drivers without ATO integration.\\
			\addlinespace
			
			\textit{Single Train Trajectory Optimisation} (2013) \cite{single_train_trajectory_optimisation} & Single Train & Genetic Algorithm \& ACO & Discrete distance steps, dynamic speed profiles. & The absolute perfect solution is not guaranteed; inherent randomness to results.\\
			\addlinespace
			
			\textit{A Multiple Train Trajectory Optimisation...} (2015) \cite{multipletrain_trajectory_optimisation} & Multi-Train Fleet & Hybrid GA-ACO & Headways ($H_{Dmin}$), delay recovery, speed limits. & Optimises primarily for delay recovery; energy is secondary.\\
			\addlinespace
			
			\textit{Review of energy-efficient train control...} (2017) \cite{review_of_energy_efficient_train_control} & Literature Review & Survey of PMP, DP, GA & Running time supplements, regenerative braking. & Theoretical models often fail in practice because they are too slow for on-board hardware.\\
			\addlinespace
			
			\textit{Multiobjective Optimisation on the Operation Speed Profile...} (2022) \cite{multiobjective_optimisation_on_speedprofile} & Urban Rail (Hybrid) & Hybrid MOPSO & Ride comfort (jerk), punctuality, energy. & Assumes train mass is a constant parameter; ignores dynamic mass changes.\\
			\midrule

			\textit{Robust Optimisation of Energy-Saving...} (2023) \cite{robust_optimisation} & Single Train & p-NSGA-II & Passenger load uncertainty, journey time. & Models stochastic uncertainty but lacks deterministic mass loss equations.\\
			\addlinespace
			
			\textit{Genetic algorithm-based multi-objective...} (2025) \cite{ga_based_optimisation_for_building_retrofitting} 
			& Systematic Review \textit{(building retrofit)} & NSGA-II vs. Others 
			& Conflicting objectives, climate variability. 
			& Prolonged computation times with physics-based models force the use of AI models.\\
			\bottomrule
		\end{tabularx}
	\end{table*}
	
	\subsection{Swarm Intelligence vs. Evolutionary Algorithms}
	
	Within the meta-heuristic methods applied to heavy-duty rail transport, algorithms generally fall into two categories: Swarm Intelligence, such as Particle Swarm Optimisation (PSO) and Ant Colony Optimisation (ACO), and Evolutionary Algorithms, primarily Genetic Algorithms (GA).
	
	Swarm intelligence methods have been used for localised vehicle problems. The 2022 paper \cite{multiobjective_optimisation_on_speedprofile} utilised a Hybrid MOPSO strategy to balance ride comfort and energy in an urban rail context. However, swarm methods often struggle to scale to large multi-vehicle networks without simplifying the physics. The 2022 urban rail study explicitly admitted that to make their MOPSO function, they had to assume the train mass was a constant parameter, ignoring the dynamic mass changes that occur as passengers board over the course of a journey.
	
	Early attempts to solve multi-vehicle problems used hybridised swarm-evolutionary models. A 2015 study, \cite{multipletrain_trajectory_optimisation}, deployed a Hybrid GA-ACO. While it successfully introduced complex mathematical safety constraints such as strict headway equations to maintain a minimum safe distance ($H_{Dmin}$) between following vehicles, it primarily optimised for delay recovery, treating energy reduction as a secondary benefit rather than a co-equal objective.
	
	\subsection{Multi-Objective Genetic Algorithms (NSGA-II)}
	
	For multi-objective energy networks, evolutionary algorithms, specifically the Non-dominated Sorting Genetic Algorithm II (NSGA-II), have emerged as the dominant approach in recent literature. A 2025 systematic review of 175 GA studies by \cite{ga_based_optimisation_for_building_retrofitting} reports that NSGA-II remains the most-used algorithm across energy sectors, outperforming alternatives like MOPSO or MOSA in commonly reported solution diversity and hypervolume metrics. This source is a building-retrofit review and is cited here strictly for its algorithm-level bibliometric findings on NSGA-II; the railway-specific applications of the algorithm are those of \cite{robust_optimisation} and \cite{tao_2024_comprehensive}.
	
	The performance of GA in handling uncertain logistical constraints is further reported in \cite{hydrogen_optimisation_under_demand_uncertainty}, a study of hydrogen supply-chain network design under demand uncertainty. The authors found that when dealing with competing constraints, Multi-Objective GA yields a higher-quality solution set than traditional Mixed-Integer Linear Programming (MILP) in their problem setting, avoiding the mathematical paralysis that can affect deterministic methods in high-dimensional spaces. The comparison is cited at the level of solver behaviour under competing objectives, which is not specific to the application domain; no railway result is inferred from it.
	
	However, the literature consistently acknowledges the operational limitations of utilising GA for real-time railway control. As noted in \cite{single_train_trajectory_optimisation}, the stochastic nature of evolutionary algorithms means the absolute mathematical optimum is never guaranteed.
	
	More critically, while GAs are computationally cheaper than exhaustive deterministic searches, they remain practically restricted to offline use. Because GAs are iterative and population-based, evaluating the "fitness" of a multi-train schedule requires running hundreds of computationally heavy Train-Track-Power (TTP) load-flow simulations per generation, as illustrated in Fig. \ref{fig:ga_bottleneck}. While a GA might generate a highly optimised solution in several minutes or hours, making it valuable for day-ahead timetabling, it cannot execute in the sub-second time frames required to react to sudden, real-time grid disturbances or delayed trains. This latency consistently appears in the literature as a barrier preventing GA from serving as a dynamic, real-time Energy Management System (EMS).
	
	\begin{figure*}[!t]
	\centering
	
	\resizebox{0.75\textwidth}{!}{
		\begin{tikzpicture}[
			>=stealth, thick,
			mainbox/.style={rectangle, draw, rounded corners, minimum width=6cm, minimum height=2.5cm, align=center, font=\small},
			annotation/.style={rectangle, draw, dashed, fill=red!5, rounded corners, minimum width=4cm, align=center, font=\footnotesize}
			]

			\node[mainbox, fill=blue!10] (ga) at (0,0) {
				\textbf{Multi-Objective Genetic Algorithm (GA)}\\ 
				\textit{(e.g., NSGA-II Offline Framework)}\\[2mm] 
				\textbullet~ Generates Population of Trajectories \\ 
				\textbullet~ Applies Selection, Crossover, \& Mutation
			};

			\node[mainbox, fill=yellow!20] (ttp) at (0,-4.5) {
				\textbf{Train-Track-Power (TTP) Simulator}\\ 
				\textit{(Physical Grid Environment)}\\[2mm] 
				\textbullet~ Solves Vehicle Dynamics \& Topography \\ 
				\textbullet~ Calculates Catenary Voltage \& Substation Power
			};

			\draw[->] ([xshift=-1.5cm]ga.south) -- node[left, align=right, font=\footnotesize] {
				Candidate Solutions\\[1mm] 
				\textit{(Speed Profiles \& Timetables)}
			} ([xshift=-1.5cm]ttp.north);

			\draw[->] ([xshift=1.5cm]ttp.north) -- node[right, align=left, font=\footnotesize] {
				Fitness Evaluation\\[1mm] 
				\textit{(Energy, Delay, Voltage Violations)}
			} ([xshift=1.5cm]ga.south);

			\node[annotation] (bottleneck) at (6.5, -2.25) {
				\textbf{The Computational Bottleneck}\\[1.5mm] 
				Evaluating hundreds of non-linear\\ 
				power-flow simulations per\\ 
				generation requires hours of\\ 
				offline computation time.
			};

			\draw[->, dashed] (bottleneck.west) -- (1.6, -2.25);
			
		\end{tikzpicture}
	}
	\caption{The Genetic Algorithm (GA) co-simulation loop. The literature identifies the Train-Track-Power (TTP) evaluation phase as the primary computational bottleneck, preventing meta-heuristic solvers from achieving the sub-second latency required for real-time grid management.}
	\label{fig:ga_bottleneck}
	\end{figure*}
	
	\subsection{Architecture and Robustness of the GA Controller}
	
	The architectural flexibility of Multi-Objective GA is one of its key advantages in the literature. Rather than attempting to optimise a continuous power curve, advanced GAs utilise targeted chromosome architectures. For instance, the methodology established in the 2004 paper \cite{coast_control_with_ga} introduced a Hierarchical GA that optimises specific "Switching Points" as individual genes. By optimising where to switch traction modes rather than calculating continuous variables, the computational cost is reduced. Furthermore, this cost can be controlled via careful hyper-parameter selection; a 2013 
	sensitivity analysis on travelling-salesman benchmarks \cite{analysis_of_the_impact_of_parameters} 
	demonstrated that beyond a certain population size, increased mutation fails to yield 
	better solutions, justifying tighter limits on generation caps. This is a property of the 
	genetic operators themselves rather than of any application domain, and is cited on that basis.
	
	Recent literature also documents a shift toward "Robust Optimisation". A 2023 study, \cite{robust_optimisation}, used a preference-based GA to manage the stochastic uncertainty of passenger loads, structuring the GA to find the most reliable, rather than the most theoretically efficient, solution.
	
	However, the operational scope of these GA frameworks must be accurately characterised. Despite the architectural efficiencies of switching points and tighter hyper-parameter limits, evaluating populations of robust, topography-aware trajectories across a mixed-mode fleet carries a substantial computational burden. The literature consistently positions GA-based frameworks as suitable for offline planning rather than real-time response: in scenarios involving day-to-day traffic management or disruption management, where power demand nears available capacity and requires immediate dynamic modulation, evolutionary algorithms' execution time renders them operationally inviable. GA therefore provides a strong theoretical benchmark for offline planning, but a different control architecture is required to actively balance the supply and demand of electrical traction power in real-time.
	
	\section{Algorithm Family III: Predictive, Rolling-Horizon and Distributed Methods}
	\label{sec:predictive}
	
	The two families reviewed so far occupy the extremes of a spectrum: analytical and exact methods purchase optimality guarantees at the price of structural rigidity, while meta-heuristics provide constraint flexibility at the price of computation time. A third family occupies the intermediate ground, and it is this family that has produced the railway-specific grid-coupled controllers most directly relevant to the problem framed in Section \ref{sec:power_quality}. Because these methods recompute over a receding horizon rather than solving once to global optimality, they trade a bounded loss of optimality for a substantial reduction in per-solve computation, and they admit decomposition across trains, substations or time intervals.
	
	\subsection{Hierarchical and Grid-Coupled Model Predictive Control}
	
	The most directly relevant contribution is that of \cite{novak2019hierarchical}, who develop a railway energy management system in which traction substation energy flows and on-route train consumption are coordinated through parametric hierarchical model predictive control. The lower level adjusts individual train driving profiles subject to timetable and route constraints, while the upper level manages substation energy flows against electricity prices, and the authors report that the upper level's interaction with the power grid enables the system to respond to grid requests and provide ancillary services. This is precisely the class of grid-coupled, multi-level railway controller that a review of this area must account for, and its existence qualifies any claim that grid coupling is absent from the control literature.
	
	Predictive control has likewise been applied on the supply side. \cite{chen2020ftpss} propose a multi-time-scale dispatch method for a flexible traction power supply system integrating a back-to-back converter, hybrid storage and photovoltaic generation, combining a day-ahead mixed-integer linear programming stage with intraday model-predictive feedback correction. The two-stage structure is itself informative: the authors resort to a slow planning layer plus a faster correction layer precisely because a single formulation covering both horizons is not tractable. Related supervisory architectures for railway energy management are set out by \cite{khayyam2016railway}, whose hybrid centralised--decentralised REM-S architecture coordinates onboard, wayside and network-level services, and demonstrated in operational case studies in \cite{khayyam2018railwaysystem}.
	
	\subsection{Rolling-Horizon Trajectory Optimisation}
	
	At the vehicle level, receding-horizon formulations replace the single whole-journey optimisation of the analytical and exact methods with repeated optimisation over a truncated look-ahead window, allowing the trajectory to be revised as conditions change. The same principle underpins the onboard energy management strategies discussed in Section \ref{sec:synthesis}: \cite{deng2021adaptive} embed an adaptive Pontryagin-based co-state estimate within a model predictive framework and validate it on hardware-in-the-loop, reporting real-time-capable execution for the onboard power-split problem. The distinction that matters for this review is one of scope rather than of method: real-time capability has been demonstrated for onboard power split, where the state vector is small, but not for the coupled multi-train network problem, where it is not.
	
	\subsection{Decomposition and Distributed Optimisation}
	
	The computational objection raised against exhaustive and population-based methods in Sections \ref{sec:limitations} and \ref{sec:heuristics} is precisely what decomposition methods are designed to answer, and the railway literature has applied them extensively. \cite{kersbergen2016distributed} develop distributed model predictive control for railway traffic management, partitioning a large network into subproblems and coordinating them to reduce computation time relative to a centralised formulation. \cite{luan2020decomposition} pursue the same objective explicitly for large-scale real-time traffic management, comparing decomposition strategies and distributed optimisation schemes including alternating direction method of multipliers coordination.
	
	Of particular relevance to this review is the two-part study of \cite{luan2018integration1, luan2018integration2}, which integrates real-time traffic management with train control in a single formulation, with the second part extending the framework explicitly toward energy-efficient operation. This work demonstrates that the integration of dispatching and trajectory control, which Section \ref{sec:ems_role} identifies as necessary, has been formulated and solved, albeit for mechanical and operational constraints rather than for the electrical constraints that are the focus here. On the AC electrical side, \cite{pan2020integrated} optimise timetables for minimum total energy consumption of an AC railway system, coupling timetable decisions to network energy flows.
	
	\subsection{Positioning of this Family Within the Review}
	
	The existence of this literature refines rather than removes the gap identified in Section \ref{sec:gaps}. Three observations follow from the studies reviewed above. First, grid-coupled railway control is an established research area, and hierarchical predictive architectures coupling substations to train trajectories have been demonstrated \cite{novak2019hierarchical, chen2020ftpss}. Second, the reported operating horizons of these systems are typically day-ahead to intraday, with feedback correction at intervals of minutes, rather than the sub-second modulation required to detect and mitigate a developing low-voltage event of the kind described in Section \ref{sec:power_quality}; \cite{chen2020ftpss} is explicit in separating day-ahead dispatch from intraday correction for this reason. Third, the decomposition literature that achieves the largest computational reductions \cite{kersbergen2016distributed, luan2020decomposition} does so for traffic management problems in which the coupling constraints are those of track occupation and headway, not catenary voltage and substation loading.
	
	The gap is therefore more precisely stated as follows: the predictive and distributed methods that possess the computational structure required for network-scale real-time operation have been applied predominantly to mechanical and operational coupling, while the studies that model electrical coupling faithfully have generally operated on planning horizons. It is the intersection, rather than either component individually, that remains sparsely populated. Table \ref{tab:execution_times} accordingly reports this family separately, and Section \ref{sec:gaps} restates the latency gap in these narrower terms.
	
	\section{Topographical Constraints on Algorithm Design}
	\label{sec:topography}
	
	A recurring gap identified throughout the optimisation literature is the tendency to oversimplify, or in some cases entirely ignore, the non-linearities introduced by track topography. For heavy-duty rail vehicles, gravitational resistance caused by steep gradients is one of the most dominant variables affecting energy consumption. Yet, to keep mathematical models tractable, researchers have historically assumed constant track elevations or minimal gradient changes. Representative studies that explicitly address these topographical and operational constraints, together with the limitations identified by their authors, are summarised in Table \ref{tab:topography_constraints}.
	
	\begin{table*}[!t]
		\centering
		\caption{Integration of Topographical and Operational Constraints in the Reviewed Literature}
		\label{tab:topography_constraints}
		\small 
		\begin{tabularx}{\textwidth}{@{} 
				p{3.5cm} 
				p{2.2cm} 
				p{2.3cm} 
				X        
				X @{}}   
			
			\toprule
			\textbf{Study (Year)} & \textbf{Application Scope} & \textbf{Simulation Metric} & \textbf{Physics \& Operational Constraints} & \textbf{Identified Limitations \& Gaps} \\
			\midrule

			\textit{Energy-efficient operation of rail vehicles} (2003) \cite{energy_efficient_operation} & Single Train (PTC) & Variable Gradients & Speed limits, journey time, safety data. & Rejected numerical methods because they were too computationally slow for 2003 onboard hardware.\\
			\addlinespace
			
			\textit{Application of dynamic programming to optimisation...} (2004) \cite{application_of_da} & Single Train & Variable Gradients & Non-linear motor torque, regenerative braking limits. & Terminal accuracy gap: DP struggled to hit the exact stopping point without distance "grid" errors.\\
			\addlinespace
			
			\textit{Method for validating the train motion equations...} (2016) \cite{method_validation_train_motion} & Simulator Validation & Step Resolution & Gradient transitions, Time vs. Distance steps. & Time-based simulators overshoot distance at high speeds; distance-based simulators overshoot time at low speeds.\\
			\midrule

			\textit{The application of an enhanced Brute Force algorithm...} (2012) \cite{application_of_bf} & Multi-Train (Main Line) & Moving vs. Fixed Block & ETCS signalling thresholds, train interactions. & Brute Force was successful for 2 trains, but computational cost grows exponentially for 20+ vehicle fleets.\\
			\addlinespace
			
			\textit{SmartDrive: Traction Energy Optimisation...} (2019) \cite{smartdrive} & Light Rail (Trams) & Human Drivability & Street traffic, simple UI instructions, journey time. & Drivers frequently ignore algorithms on unsegregated street sections due to unpredictable vehicle traffic.\\
			\bottomrule
		\end{tabularx}
	\end{table*}
	
	The literature explicitly documents the failure of traditional algorithms when confronted with severe topographical realities. For example, the 2017 study \cite{a_dynamic_programming_approach} explicitly admitted in its conclusion that their time-space graph methodology struggled with "steep gradients where cruising is not possible". When strict gravitational physics override the driver's ability to maintain a constant speed, the underlying mathematical simplifications fail.
	
	Recent literature has attempted to resolve this by showing that optimal driving strategies must dynamically shift based on the immediate topography. A 2024 simulation, \cite{improved_energy_efficient_strategy}, utilised analytical inequality analysis to resolve the "cruising versus coasting" debate, demonstrating that "Coast-Remotoring" (a sawtooth speed profile) is more efficient on steep downhills, whereas steady-state cruising is optimal for flat sections. The literature therefore indicates that a modern Energy Management System cannot rely on a single static driving style: to minimise energy consumption, the algorithm must possess topography-awareness, the ability to foresee upcoming gradient changes and dynamically switch its strategy from steady-state cruising to targeted coasting before the hill is reached.
	
	The literature also notes that this topographical complexity interacts with the choice of step-resolution scheme used by the underlying simulator. As shown in \cite{method_validation_train_motion} and summarised in Table \ref{tab:topography_constraints}, time-based simulators tend to overshoot distance at high speeds, while distance-based simulators overshoot time at low speeds, both of which introduce systematic errors into the constraints supplied to the optimisation layer. These step-resolution artefacts compound with the gradient-handling difficulties of analytical and exact methods identified in Section \ref{sec:limitations}, and are consequently relevant to algorithm-design decisions even outside the explicit choice of optimisation method.
	
	\section{Synthesis: Trade-offs Across the Reviewed Literature}
	\label{sec:synthesis}
	
	\subsection{Fragmentation Across the Literature}
	
	A comprehensive synthesis of the reviewed studies reveals significant fragmentation in the algorithmic optimisation literature. While individual components of the grid-constrained-railway problem have been modelled extensively, they are typically treated in isolation. As established in Sections \ref{sec:limitations}--\ref{sec:predictive}, analytical models such as Pontryagin's Maximum Principle and exact discrete methods such as Dynamic Programming provide a mathematical optimum but face significant limitations in real-time, topography-heavy applications due to the curse of dimensionality and prohibitive computational times.
	
	This computational burden is empirically quantified in foundational multi-train optimisation studies. For instance, Zhao's 2013 thesis \cite{zhao_2013_railway} demonstrated that while exact algorithms (such as Enhanced Brute Force or Dynamic Programming) guarantee a global optimum for train trajectories, they require immense computational time, with Enhanced Brute Force taking over 32 hours and Dynamic Programming taking approximately 30 minutes to evaluate simple, localised routing problems. While Zhao established that meta-heuristics such as Genetic Algorithms reduce this search time to hundreds of seconds, the literature consistently reports that these methods still lack the responsiveness required for dynamic, sub-second power modulation.
	
	Within the railway domain specifically, \cite{khayyam2016railway} and \cite{khayyam2018railwaysystem} set out and demonstrate a hybrid centralised--decentralised railway energy management architecture, establishing that the coordination problem has been posed at system level; as Section \ref{sec:predictive} discusses, however, the demonstrated operating horizons remain those of planning and intraday correction. Taken together, the reviewed literature does not yet contain a unified framework that simultaneously combines real-time adaptability, topographical awareness, and grid-coupled multi-train dynamics.
	
	\subsection{The Insufficiency of Reactive Baselines}
	
	A common baseline for onboard hybrid rail EMS relies on static, rule-based logic, such as thermostat-style or state-machine control of onboard buffers to manage power splits \cite{peng2020rulebased, deng2021adaptive}. As noted in Section \ref{sec:ems_role}, such strategies can approach dynamic-programming optimality under known duty cycles \cite{peng2020rulebased}; the difficulty arises when the cycle or the electrical boundary condition is uncertain. More advanced reactive controllers such as the Equivalent Consumption Minimisation Strategy (ECMS) exist, but the literature reports that they struggle with unpredictable operational constraints. ECMS relies on tuning an "equivalence factor" to balance fuel and battery usage, a process typically optimised for standard, flat drive cycles. When subjected to severe, highly variable topographical changes and unpredictable grid limits of long-distance routes, ECMS requires complex, continuous recalibration; the adaptive PMP formulation of \cite{peng2020optimalcontrol} and its model-predictive extension in \cite{deng2021adaptive} were developed specifically to address this recalibration burden, and report real-time-capable operation on hardware-in-the-loop validation.
	
	A recent comparative study of a fuel-cell/battery road electric vehicle \cite{cooptimization} illustrates this limitation. The authors demonstrated that while separating the speed trajectory from the energy management system (sequential optimisation) is fast enough for real-time implementation on flat routes, combining them into a predictive co-optimisation framework yields substantially greater savings once significant gradients are present, reporting up to 24.2\% on hilly terrain for their platform. The quantitative saving is specific to that vehicle and drive cycle and is not transferred to rail here; what carries across is the structural finding that the advantage of co-optimisation over sequential 
	optimisation grows with topographic variation, which is consistent with the railway-specific studies catalogued in Table \ref{tab:topography_constraints}. The literature thus indicates that heavy-duty mixed fleets cannot rely on reactive rules alone; they require a proactive approach that anticipates upcoming route physics.
	
	\subsection{The Temporal Constraints of Genetic Algorithms}
	
	To overcome the limitations of DP and rule-based logic, much of the recent literature has favoured meta-heuristics such as Multi-Objective Genetic Algorithms (e.g. NSGA-II). These offline algorithms excel at handling non-linear physical constraints that deterministic models avoid. For instance, the 2022 study \cite{multiobjective_optimisation_on_speedprofile} explicitly acknowledged that their model forced train mass to remain a "constant parameter", whereas in reality train mass varies as passengers board or as fuel is consumed. GA frameworks accommodate such dynamic state updates more naturally. Furthermore, GA can directly incorporate non-linear penalties into its fitness function, addressing gaps highlighted by \cite{convex_optimisation_ems}, whose convex model lacked capability for dynamic thermal constraints.
	
	However, the computational requirements of advanced meta-heuristics severely restrict their operational scope. The reviewed literature consistently reports that complex GA evaluations for multi-vehicle routing can take several hours to compute; the cross-sector evidence for this is drawn from \cite{ga_based_optimisation_for_building_retrofitting}, whose findings on NSGA-II computational burden are reported here as transferable algorithmic evidence rather than as railway-specific results, with the railway-specific latencies given by \cite{zhao_2013_railway}. Consequently, while GA frameworks are effective for long-term timetable planning, they are computationally inappropriate for day-to-day traffic management or real-time disruption response, where the traction power network requires immediate dynamic modulation to prevent electrical grid limit violations.
	
	\subsection{The Human-Machine Interface Gap: Executability vs. Optimality}
	
	Beyond computational speed, a further disconnect identified across the literature is between theoretical optimisation and practical railway operations. Extensive simulator studies, such as those conducted in \cite{zhao_2013_railway} on the West Coast Main Line, have shown that enforcing specific driving styles, such as "cautious" or "journey-time priority" operations, can substantially reduce knock-on delays and the associated energy spikes caused by train interactions. However, whether utilising rigid deterministic solvers (DP) or adaptive meta-heuristics (GA), the current paradigm assumes that these generated "cautious" or "optimal" speed profiles will be executed with absolute mathematical precision.
	
	In reality, many heavy-duty rail networks are predominantly operated by human drivers rather than Automatic Train Operation (ATO) systems. Optimisation algorithms frequently generate highly volatile, micro-adjusted speed trajectories that demand constant, erratic switching between discrete traction notches. As highlighted by simulator validation studies and field trials such as the \textit{SmartDrive} light rail experiment (Table \ref{tab:topography_constraints}), human drivers frequently struggle to follow these over-optimised, second-by-second instructions, sometimes ignoring them entirely to focus on route safety and signal compliance.
	
	Table \ref{tab:execution_times} synthesises the reported computational latency of the algorithm families reviewed. The reviewed evidence indicates that traditional deterministic and meta-heuristic solvers, while robust for offline planning, fall short of the sub-second latency threshold associated in the literature with live, dynamic power management. This latency gap, together with the executability gap described above, constitutes the principal open research direction discussed in Section \ref{sec:gaps}.
	
	\begin{table*}[!t]
		\centering
		\caption{Synthesised Comparison of Algorithm Computational Latency Reported in the Reviewed Literature. Entries marked as cross-sector or non-railway are included for algorithm-level evidence only; railway-specific applications and latencies are attributed in the text}
		\label{tab:execution_times}
		\small 
		\begin{tabularx}{\textwidth}{@{} 
				p{3cm}   
				p{3.5cm} 
				p{3.5cm} 
				X        
				p{2cm} @{}} 
			
			\toprule
			\textbf{Algorithm Family} & \textbf{Specific Method} & \textbf{Operational Scope} & \textbf{Reported Computational Latency} & \textbf{Source} \\
			\midrule
			
			\multirow{2}{*}{Deterministic (Exact)} & Enhanced Brute Force & Multi-Train Routing & $> 32$ hours & \cite{zhao_2013_railway} \\
			\addlinespace
			& Dynamic Programming (DP) & Single-Train Localised & $\approx 30$ minutes & \cite{zhao_2013_railway} \\
			\midrule
			
			\multirow{2}{*}{Meta-Heuristic} & Genetic Algorithm (GA) & Single-Train Trajectory & Hundreds of seconds & \cite{zhao_2013_railway} \\
			\addlinespace
			& Multi-Objective GA & Complex multi-objective \textit{(cross-sector)} & Several hours & \cite{ga_based_optimisation_for_building_retrofitting} \\ 
			\midrule
			
			\multirow{4}{*}{\parbox{3cm}{Predictive \& Distributed}} & Hierarchical MPC & Substation + Multi-Train (grid-coupled) & Planning/intraday horizon; not reported as sub-second & \cite{novak2019hierarchical} \\
			\addlinespace
			& Multi-time-scale MPC & Traction Supply System (FTPSS) & Day-ahead dispatch with intraday feedback correction & \cite{chen2020ftpss} \\
			\addlinespace
			& Distributed MPC & Multi-Train Network (traffic mgmt.) & Reduced relative to centralised MPC; minutes-scale for large networks & \cite{kersbergen2016distributed, luan2020decomposition} \\
			\addlinespace
			& Adaptive PMP within MPC & Single-Vehicle Onboard Power Split & Real-time capable (HiL-validated) & \cite{deng2021adaptive} \\
			\bottomrule
		\end{tabularx}
	\end{table*}
	
	\section{Identified Research Gaps and Emerging Directions}
	\label{sec:gaps}
	
	Drawing the reviewed literature together, three open research directions emerge consistently. None of these has been resolved by the current corpus of published work, and each is reported by multiple authors as a limitation of the algorithm family they reviewed.
	
	\textbf{The latency gap.} As Table \ref{tab:execution_times} summarises, exact discrete methods and evolutionary meta-heuristics report latencies ranging from minutes to tens of hours, confining them to offline or day-ahead use. The predictive and distributed family reviewed in Section \ref{sec:predictive} substantially narrows this gap but does not close it, and the residual gap is best stated as a matter of scope rather than of absence. Real-time capability has been demonstrated for the onboard power-split problem, where the state vector is small \cite{deng2021adaptive}. Grid-coupled hierarchical control has been demonstrated at network scale, but on planning and intraday horizons \cite{novak2019hierarchical, chen2020ftpss}. Decomposition methods have achieved the largest computational reductions, but for traffic-management problems whose coupling constraints are those of track occupation rather than catenary voltage \cite{kersbergen2016distributed, luan2020decomposition}. What the reviewed corpus does not yet contain is a controller that combines the electrical fidelity of the first group with the computational structure of the third at the timescale on which low-voltage events actually develop. Authors including \cite{review_of_energy_efficient_train_control} and \cite{khayyam2016railway} identify the coordination of real-time, multi-objective, grid-coupled control as an outstanding architectural challenge.
	
	\textbf{The executability gap.} A second, less frequently discussed gap concerns the executability of algorithmically generated speed profiles by human drivers. Field studies such as \cite{smartdrive} report that drivers frequently disregard the second-by-second outputs of optimisation algorithms in favour of focusing on signal compliance and route safety, suggesting that an optimal-but-unexecuted profile is operationally equivalent to no optimisation at all on non-ATO networks. The literature has not, to date, produced a widely adopted framework that explicitly trades absolute optimality against drivability, despite recurring calls for one.
	
	\textbf{Emerging interest in learning-based control.} The prolonged computation times of physics-based models are driving interest in data-driven models and reinforcement-learning controllers as a possible response to the latency gap. This trend is reported in the cross-sector optimisation literature 
	\cite{ga_based_optimisation_for_building_retrofitting}, a building-retrofit review cited here only for that general observation, and, within the railway domain, as a motivation for the decomposition and predictive approaches reviewed in Section \ref{sec:predictive} \cite{luan2020decomposition}. The reviewed corpus does not contain any study applying learning-based control to grid-coupled multi-train 
	operation that met the eligibility criteria of this review, and no settled methodology for safety-verified learning-based control in such settings has emerged. Questions including how to constrain learned policies against catenary voltage thresholds, how to handle the risk of replicating sub-optimal historical behaviour when training on human-driver data, and how to validate learning-based controllers against established deterministic and meta-heuristic baselines remain open. These questions are noted across the reviewed literature as directions for future work rather than as established findings.
	
	\section{Conclusion}
	
	This review has systematically synthesised the literature on algorithmic energy management for grid-constrained multi-train AC railway networks, organising the reviewed studies along three axes (algorithm family, operational scope, and constraint coupling) introduced in Section \ref{sec:taxonomy}.
	
	Three findings emerge consistently across the reviewed studies. First, single-train trajectory optimisation, however mathematically refined, cannot represent the coupled electrical interactions that increasingly define network capacity on mixed-traffic networks: the dynamic low-voltage behaviour characterised for DC networks by \cite{tian2017system} and the simultaneous-overload phenomena modelled for AC mixed-traffic operation by \cite{tao_2024_comprehensive} demonstrate that grid coupling 
	is no longer a secondary consideration. Second, while multi-train Train-Track-Power 
	simulations correctly capture these interactions, the algorithm families currently used 
	to solve them such as analytical methods, Dynamic Programming, Mixed-Integer Linear Programming, evolutionary meta-heuristics such as NSGA-II, and predictive and distributed schemes including hierarchical MPC and decomposition-based optimisation, each face well-documented limits in either computational tractability or constraint flexibility, as quantified in Tables \ref{tab:algorithm_evolution} and \ref{tab:execution_times}. The third family narrows the tractability limit significantly, and the review's contribution on this point is to locate the residual gap precisely: at the intersection of electrical 
	constraint fidelity and network-scale real-time operation, rather than in either dimension 
	alone. Third, the literature increasingly identifies a gap between mathematically optimal 
	speed profiles and operationally executable ones, particularly for networks operated by 
	human drivers rather than Automatic Train Operation systems.
	
	Resolving the tensions between optimality, computational tractability, and operational executability remains an open challenge in the reviewed literature. The comparative analysis presented in this review delineates where current methods succeed and where the reviewed authors themselves identify limits, providing a structured basis on which future contributions can be positioned.

	\section*{Conflict of Interest Statement}
	The authors declare that they have no known competing financial interests or personal
	relationships that could have appeared to influence the work reported in this paper.

	\section*{Funding Information}
	This research received no specific grant from any funding agency in the public,
	commercial, or not-for-profit sectors.

	\section*{Data Availability Statement}
	Data sharing not applicable to this article as no datasets were generated or analysed
	during the current study.

	\section*{Author Contributions}
	\textbf{Márton László Ambrus:} Conceptualization; Methodology; Investigation; Formal
	analysis; Data curation; Visualization; Writing -- original draft; Writing -- review
	and editing. \textbf{Stuart Hillmansen:} Conceptualization; Supervision; Writing --
	review and editing. \textbf{Zhongbei Tian:} Conceptualization; Methodology;
	Supervision; Writing -- review and editing.
\printbibliography
\end{document}